\documentclass[12pt]{iopart}

\usepackage{iopams}
\usepackage{graphicx}  
\usepackage{bm}
\usepackage[hypertex]{hyperref}

\begin{document}

\title{The effect of permanent dipole moment on the polar molecule cavity quantum electrodynamics}
\author{Jing-Yun Zhao$^{1,2,3}$, Li-Guo Qin$^2$, Xun-Ming Cai$^1$, Qiang Lin$^{1,4}$, and Zhong-Yang Wang$^{2,*}$}

\address{$^1$Institute of Optics, Department of Physics, Zhejiang University, Hangzhou, 310027, China}
\address{$^2$Shanghai Advanced Research Institute, Chinese Academy of Sciences, Shanghai, 201210, China}
\address{$^3$School of Science, Zhejiang Sci.-Tech. University, Hangzhou 310018, China}
\address{$^4$Center for Optics and Optoelectronics Research, Department of Applied Physics, College of Science, Zhejiang University of Technology, Hangzhou 310023, China}
\address{$^*$Corresponding author: wangzy@sari.ac.cn}

\begin{abstract}
A dressed-state perturbation theory beyond the rotating wave approximation (RWA) is presented to investigate the interaction between a two level electronic transition of the polar molecules and a quantized cavity field. Analytical expressions can be explicitly derived for both the ground- and excited-state-energy spectrums and wave functions of the system, where the contribution of permanent dipole moments (PDM) and the counter-rotating wave term (CRT) can be shown separately. The validity of these explicit results is discussed by comparing with the direct numerical simulation. Comparing to CRT coupling, PDM results in the coupling of more dressed states and the energy shift proportional to the square of the normalized permanent dipole difference, and a greater Bloch-Siegert shift could be produced in giant dipole molecule cavity QED. In addition, our method could also be extended to the solution of two-level atom Rabi model Hamiltonian beyond the RWA.
\end{abstract}



\section{Introduction}
Polar molecules bearing permanent electric dipole moment (PDM) can provide the strong, long-range, tunable, spatially anisotropic dipole-dipole interactions \cite{1ava}. They have potential applications from high precision metrology \cite{2jjh} to chemical dynamics \cite{3ct}, quantum information processing \cite{5cmt} and the simulation of quantum many-body systems \cite{6dww} etc. The PDM reflects direct information about the electronic charge-density distribution of polar molecules, and could be quite different between various electronic states. The difference of PDM arises from the change of molecular orbital \cite{7ara,7tl,7kkd}. For instance, in the alkaline earth -monohalides and -monohydrides, the decrease of PDM from $X^2\Sigma$ state to $A^2\Pi$ state is the result of the promotion of the sole unpaired valence electron from an s-p hybrid orbital to a more highly polarizable p-d hybrid orbital \cite{7ara,7tl}. While in the alkaline monohydrides LiH, it is the result of the promotion from an s orbital to a s-p hybrid orbital \cite{7kkd}. Another type of giant dipole molecules are organic molecules with the delocalized conjugated $\pi$ electrons. Such molecules often have excited electronic states of a charge transfer character, so there can be large differences of PDM between the electronic ground and excited states \cite{shnc}. The interaction of a two-level polar molecule with a classical optical field has been studied by many researchers \cite{7mwj}. It is shown that the nonzero difference of PDM between the electronic ground and excited states can significantly affect the coupling between polar molecule and the optical field, and is responsible for some new optical properties which are quite different from a regular atomic system, for instance, multi-photon transition \cite{8tgf}, nonlinear optical properties \cite{9bn}, high harmonic generation \cite{12dla} etc.

On the other hand, the study of strong coupling between the matter and the cavity field through quantum electrodynamics (QED), in which the interaction strength exceeds the optical transition and cavity mode damping rates, has been the subject of great interests \cite{13pr}. It has broad applications ranging from metrology \cite{14tak} and quantum computation \cite{15pt} to mechanism for cooling atom and molecule \cite{18vv} etc. Nowadays, the researchers have achieved the strong coupling in different physical systems, such as atomic system \cite{19rjm}, semiconductor quantum dot microcavity \cite{20rjp}, superconducting qubit circuit QED \cite{23pf}, and organic microcavity \cite{jrt}. Also, the possibility of cavity-assisted laser cooling of polar molecules has been theoretically proposed \cite{GM}. However, in above-mentioned cavity QED interactions, little attention has been paid to the effect of permanent dipole difference between the ground state and excited state of the molecules which usually plays a role in the nonlinear process. That's our motivation to study the effect of permanent dipole difference in the field of strong coupled polar molecule cavity QED.

For the early work on strong coupled cavity QED, the coupling strength $\lambda$ was much smaller than the cavity field frequency $\omega_c$ (typical values for atom-cavity QED experiments are $\lambda/\omega_c\sim10^{-5}$ \cite{22jmr}). The interaction can therefore be well described by the Jaynes-Cummings (JC) model within the rotating-wave approximation (RWA). But recent progress draw our attention to the ultrastrong-coupling regime, for example, the normalized coupling strength $\lambda/\omega_c$ for polyelec-trolyte/J aggregate dye coupled with an optical microcavity has reached 0.1 \cite{jrt}. In this case, the counter-rotating wave terms (CRT) of the interaction can not be neglected and result in Bloch-Siegert (B-S) shift of the resonant frequency, which has been observed recently in the experiment \cite{23pf}. Therefore, the interaction of matter and cavity field is then governed by the Rabi model beyond the RWA. Because of the difficulties to obtain the exact solution of Rabi model, some approximate analytical and numerical approaches have been proposed and the B-S shift in atomic and superconducting qubit cavity QED have been studied \cite{jh,ly}.

For the strong coupled polar molecule cavity QED system, the contribution of PDM should be considered on the base of CRT \cite{th}. Because the JC-type Hamiltonian can be solved analytically in a closed form in the basis of $|e,n\rangle$ and $|g,n+1\rangle$ (where the qubit states $|e\rangle$ and $|g\rangle$ are excited state and ground state respectively, and the oscillator states $|n\rangle$ are Fock states) in terms of dressed-states \cite{26etj}, we can treat the contribution of PDM and CRT as a perturbation of interaction. In our work, a dressed-state perturbation theory (DSP) beyond the RWA is proposed to treat analytically a two-level polar molecule coupled to a quantized cavity field. Based on our DSP theory, the B-S shift in the polar molecule cavity QED system is also obtained, which includes the contribution of both CRT and PDM separately. The B-S shift caused by the CRT is approximately proportional to the square of the coupling strength, while the B-S shift caused by the PDM is proportional to the square of the normalized permanent dipole difference and dependent on the higher terms of $\lambda$, which is coincident with Ref. \cite{th}.  In addition, for the transition between the higher excited state and the ground state, the magnitude of the BS shift becomes larger as energy level increases. Also, Our method could be extended to the solution of two-level atom Rabi model Hamiltonian beyond the RWA by setting normalized permanent dipole difference equals zero.

The paper is organized as follows. In Sec. 2, we present the DSP theory and derive analytical expressions for the eigenenergies and eigenstates of a two-level polar molecule cavity QED system. The validity of our approach up to second-order correction is also given. In Sec. 3, we show the transition relation between different energy levels through their populations, and the respective contribution of PDM and CRT to the transition is given as well. In addition, by comparing the energy spectrum obtained by our approach with those by numerical simulation, we furthermore clarify the validity of our method. The respective effect of PDM and CRT on the energy level shift are also analyzed. Finally, conclusions are drawn in Sec. 4.

\section{Analytical expressions for the eigenenergies and eigenstates}

In this paper, we focus on the coupling of electronic transition (such as alkaline earth monohalides \cite{7ara}, alkaline earth monohydrides \cite{7tl}, alkaline monohydrides \cite{7kkd}, giant dipole diphenyl molecules \cite{shnc} etc.) with the visible laser field. As for the coupling of vibrational transition or rotational transition of polar molecule with cavity field, which are not dealt with here, they will be discussed in our future work. When the frequency of the relevant cavity mode $\omega_c$ is near resonant with a polar molecule's transition frequency $\omega_0$ between electronic ground state and excited state, the dynamics of the molecule is restricted to the two state $\left\{|g\rangle,|e\rangle\right\}$, and the molecule is well approximated by a two-level system.

The interaction between a two-level polar molecule and a quantized cavity field with frequency $\omega_c$ is described by \cite{th}:
\begin{equation}
H=\omega_ca^{\dag}a+\frac{1}{2}\omega_0\sigma_z-\lambda(\alpha\sigma_z+\sigma_x)(a^{\dag}+a),
\end{equation}
where $a^{\dag}$ and $a$ are the creation and annihilation operators of a photon, $\sigma_x$ and $\sigma_z$ are Pauli spin operators, and we have set $\hbar$ to unity. The ground state and excited state of the polar molecule have permanent dipole moments $\mu_{gg}$ and $\mu_{ee}$, respectively. $\mu_{ge}$ is the transition dipole moment. $ 2\alpha=\frac{\mu_{ee}-\mu_{gg}}{\mu_{ge}} $ is the normalized permanent dipole difference. $\lambda=\mu_{ge}\sqrt{\frac{\omega_c}{2\varepsilon_0V}}$ is the coupling strength, where $V$ is the cavity mode volume. The specific parameters of permanent dipole moment $\mu_{gg}$ and $\mu_{ee}$, transition dipole moment $\mu_{ge}$, and transition frequency $\omega_0$ are given in the specific molecule references.

We consider the Hamiltonian can be divided into two parts:
\begin{equation}
H=H_0+H^{\prime},
\end{equation}
consisting of
\begin{equation}
H_0=\omega_ca^{\dag}a+\frac{1}{2}\omega_0\sigma_z-\lambda(a\sigma_{+}+a^{\dag}\sigma_{-})
\end{equation}
and
\begin{equation}
H^{\prime}=-\lambda[\alpha\sigma_z(a^{\dag}+a)+\sigma_{+}a^{\dag}+\sigma_{-}a],
\end{equation}
where the first part $H_0$ is the JC model Hamiltonian within the RWA, whose eigenvalues and eigenstates are exactly solvable in terms of dressed-states, and the second part $H^{\prime}$ involves the interaction Hamiltonian of both the PDM and the CRT. If $H^{\prime}$  is a very small quantity compared with $H_0$, then we can treat $H^{\prime}$ as a perturbation part, and derive approximated solution of eigenvalue and eigenfunction of $H$. The validity of the DSP theory is discussed later in the paper.

First, we can obtain energy eigenstates of the unperturbed system through the diagonalization of non-perturbation part Hamiltonian $H_0|\varphi_n^{\pm}\rangle=\varepsilon_n^{\pm}|\varphi_n^{\pm}\rangle$. $|\varphi_n^{+}\rangle=\sin\theta_n|e,n\rangle+\cos\theta_n|g,n+1\rangle$, $|\varphi_n^{-}\rangle=\cos\theta_n|e,n\rangle-\sin\theta_n|g,n+1\rangle$, where $\tan\theta_n=\frac{2\lambda\sqrt{n+1}}{\delta-\Omega_{n\delta}}$, detuning $\delta=\omega_0-\omega_c$, Rabi frequency $\Omega_{n\delta}=\sqrt{\delta^2+4\lambda^2(n+1)}$. The corresponding eigenenergies are: $\varepsilon_n^{\pm}=\omega_c(n+\frac{1}{2})\pm\frac{\Omega_{n\delta}}{2}$. $|g,0\rangle$ is also a eigenstate of $H_0$ with eigenenergy $-\omega_0/2$. These energy eigenstates form a orthogonal and normalized completed set: $\langle\varphi_m^{+}|\varphi_n^{+}\rangle=\langle\varphi_m^{-}|\varphi_n^{-}\rangle=\delta_{mn}$, $\langle\varphi_m^{+}|\varphi_n^{-}\rangle=\langle\varphi_m^{\pm}|g,0\rangle=0$, $\sum_{n=0}^{\infty}(|\varphi_n^{+}\rangle\langle\varphi_n^{+}|+|\varphi_n^{-}\rangle\langle\varphi_n^{-}|)+|g,0\rangle\langle g,0|=1$.

Since these unperturbed eigenstates are completed, the explicit state $|\Psi\rangle$ of $H$ can be expanded in terms of the eigenfunction set as:
\begin{equation}
|\Psi\rangle=\sum_{n=0}^{\infty}(a_n|\varphi_n^{+}\rangle+b_n|\varphi_n^{-}\rangle)+c|g,0\rangle,
\end{equation}
the detailed description of dressed-state perturbation theory is given in Appendix A.

Suppose that the system is in one of the eigenstates $|\varphi_k^{\pm}\rangle$ (or $|g,0\rangle$) before being perturbed, which means that zero-order approximated wave function is $|\Psi_k^{\pm(0)}\rangle=|\varphi_k^{\pm}\rangle$ (or $|g,0\rangle$) and zero-order approximated energy is $E_k^{\pm(0)}=\varepsilon_k^{\pm}$ (or $-\omega_0/2$). Based on the framework of perturbation theory, analytical approximations for the wave function and the energy of dressed-state after being perturbed can be performed systematically. The ground-state wave function and energy can be obtained as (all of them are up to the second-order correction):
\begin{equation}
|\Psi_{g,0}\rangle=|\Psi_{g,0}^{(0)}\rangle+|\Psi_{g,0}^{(1)}\rangle+|\Psi_{g,0}^{(2)}\rangle
\end{equation}

\begin{equation}
E_{g,0}=E_{g,0}^{(0)}+E_{g,0}^{(1)}+E_{g,0}^{(2)}
\end{equation}

where $|\Psi_{g,0}^{(0)}\rangle=|g,0\rangle$, $|\Psi_{g,0}^{(1)}\rangle=|\Psi_{g,0(\alpha)}^{(1)}\rangle+|\Psi_{g,0(CRT)}^{(1)}\rangle$, $|\Psi_{g,0}^{(2)}\rangle=|\Psi_{g,0(\alpha^2)}^{(2)}\rangle+|\Psi_{g,0(CRT)}^{(2)}\rangle+|\Psi_{g,0(\alpha,CRT)}^{(2)}\rangle$, and $E_{g,0}^{(0)}=-\frac{\omega_0}{2}$, $E_{g,0}^{(1)}=0$, $E_{g,0}^{(2)}=E_{g,0(\alpha^2)}^{(2)}+E_{g,0(CRT)}^{(2)}$. Also, the excited-state wave functions and energy spectrums can be given by:

\begin{equation}
|\Psi_k^{\pm}\rangle=|\Psi_k^{\pm(0)}\rangle+|\Psi_k^{\pm(1)}\rangle+|\Psi_k^{\pm(2)}\rangle
\end{equation}

\begin{equation}
E_k^{\pm}=E_k^{\pm(0)}+E_k^{\pm(1)}+E_k^{\pm(2)}
\end{equation}

where $|\Psi_k^{\pm(0)}\rangle=|\varphi_k^{\pm}\rangle$, $|\Psi_k^{\pm(1)}\rangle=|\Psi_{k(\alpha)}^{\pm(1)}\rangle+|\Psi_{k(CRT)}^{\pm(1)}\rangle$, $|\Psi_k^{\pm(2)}\rangle=|\Psi_{k(\alpha^2)}^{\pm(2)}\rangle+|\Psi_{k(CRT)}^{\pm(2)}\rangle+|\Psi_{k(\alpha,CRT)}^{\pm(2)}\rangle$, and $E_k^{\pm(0)}=\varepsilon_k^{\pm}$, $E_k^{\pm(1)}=0$, $E_k^{\pm(2)}=E_{k(\alpha^2)}^{\pm(2)}+E_{k(CRT)}^{\pm(2)}$,

\begin{equation}
E_{g,0(\alpha^2)}^{(2)}=-\lambda^2\alpha^2\left(\frac{cos^2\theta_0}{\frac{\omega_0}{2}+\varepsilon_0^{+}}+\frac{sin^2\theta_0}{\frac{\omega_0}{2}+\varepsilon_0^{-}}\right)
\end{equation}

\begin{equation}
E_{g,0(CRT)}^{(2)}=-\lambda^2\left(\frac{sin^2\theta_1}{\frac{\omega_0}{2}+\varepsilon_1^{+}}+\frac{cos^2\theta_1}{\frac{\omega_0}{2}+\varepsilon_1^{-}}\right)
\end{equation}

\begin{eqnarray}
E_{k(\alpha^2)}^{+(2)}=&&\lambda^2\alpha^2\bigg[\frac{(\sqrt{k+1}sin\theta_ksin\theta_{k+1}-\sqrt{k+2}cos\theta_kcos\theta_{k+1})^2}{\varepsilon_k^{+}-\varepsilon_{k+1}^{+}}
\nonumber\\&&+\frac{(\sqrt{k+1}sin\theta_kcos\theta_{k+1}+\sqrt{k+2}cos\theta_ksin\theta_{k+1})^2}{\varepsilon_k^{+}-\varepsilon_{k+1}^{-}}
\nonumber\\&&+\frac{(\sqrt{k}sin\theta_ksin\theta_{k-1}-\sqrt{k+1}cos\theta_kcos\theta_{k-1})^2}{\varepsilon_k^{+}-\varepsilon_{k-1}^{+}}
\nonumber\\&&+\frac{(\sqrt{k}sin\theta_kcos\theta_{k-1}+\sqrt{k+1}cos\theta_ksin\theta_{k-1})^2}{\varepsilon_k^{+}-\varepsilon_{k-1}^{-}}
\nonumber\\&&+\frac{1}{\varepsilon_k^{+}+\frac{\omega_0}{2}}\left(\sqrt{k+1}cos\theta_k\langle0|k\rangle\right)^2\bigg]
\end{eqnarray}

\begin{eqnarray}
E_{k(CRT)}^{+(2)}=&&\lambda^2\Bigg[(k+2)cos^2\theta_k\left(\frac{sin^2\theta_{k+2}}{\varepsilon_k^{+}-\varepsilon_{k+2}^{+}}+\frac{cos^2\theta_{k+2}}{\varepsilon_k^{+}-\varepsilon_{k+2}^{-}}\right)
\nonumber\\&&+ksin^2\theta_k\left(\frac{cos^2\theta_{k-2}}{\varepsilon_k^{+}-\varepsilon_{k-2}^{+}}+\frac{sin^2\theta_{k-2}}{\varepsilon_k^{+}-\varepsilon_{k-2}^{-}}\right)
\nonumber\\&&+\frac{1}{\varepsilon_k^{+}+\frac{\omega_0}{2}}\left(\sqrt{k}sin\theta_k\langle0|k-1\rangle\right)^2\Bigg]
\end{eqnarray}

\begin{eqnarray}
E_{k(\alpha^2)}^{-(2)}=&&\lambda^2\alpha^2\bigg[\frac{(\sqrt{k+1}cos\theta_ksin\theta_{k+1}+\sqrt{k+2}sin\theta_kcos\theta_{k+1})^2}{\varepsilon_k^{-}-\varepsilon_{k+1}^{+}}
\nonumber\\&&+\frac{(\sqrt{k+1}cos\theta_kcos\theta_{k+1}-\sqrt{k+2}sin\theta_ksin\theta_{k+1})^2}{\varepsilon_k^{-}-\varepsilon_{k+1}^{-}}
\nonumber\\&&+\frac{(\sqrt{k}sin\theta_{k-1}cos\theta_{k}+\sqrt{k+1}cos\theta_{k-1}sin\theta_{k})^2}{\varepsilon_k^{-}-\varepsilon_{k-1}^{+}}
\nonumber\\&&+\frac{(\sqrt{k}cos\theta_{k-1}cos\theta_{k}-\sqrt{k+1}sin\theta_{k-1}sin\theta_{k})^2}{\varepsilon_k^{-}-\varepsilon_{k-1}^{-}}
\nonumber\\&&+\frac{1}{\varepsilon_k^{-}+\frac{\omega_0}{2}}\left(\sqrt{k+1}sin\theta_k\langle0|k\rangle\right)^2\bigg]
\end{eqnarray}

\begin{eqnarray}
E_{k(CRT)}^{-(2)}=&&\lambda^2\Bigg[(k+2)sin^2\theta_k\left(\frac{sin^2\theta_{k+2}}{\varepsilon_k^{-}-\varepsilon_{k+2}^{+}}+\frac{cos^2\theta_{k+2}}{\varepsilon_k^{-}-\varepsilon_{k+2}^{-}}\right)
\nonumber\\&&+kcos^2\theta_k\left(\frac{cos^2\theta_{k-2}}{\varepsilon_k^{-}-\varepsilon_{k-2}^{+}}+\frac{sin^2\theta_{k-2}}{\varepsilon_k^{-}-\varepsilon_{k-2}^{-}}\right)
\nonumber\\&&+\frac{1}{\varepsilon_k^{-}+\frac{\omega_0}{2}}\left(\sqrt{k}cos\theta_k\langle0|k-1\rangle\right)^2\Bigg]
\end{eqnarray}

All of above results can be obtained by straight but complex calculation, and the detailed expression of ground state wavefunction $|\Psi_{g,0}\rangle$ is given in Appendix B. Of course, according to the similar procedure, we can also get the wave function of the higher order dressed-state $|\Psi_k^{\pm}\rangle$, but not all of them are shown here, and only the first-order wavefunction correction $|\Psi_k^{\pm(1)}\rangle$ are shown in Appendix C. In all of these analytical expressions, $|\Psi_{g,0(\alpha)}^{(1)}\rangle, |\Psi_{k(\alpha)}^{\pm(1)}\rangle, |\Psi_{g,0(\alpha^2)}^{(2)}\rangle, |\Psi_{k(\alpha^2)}^{\pm(2)}\rangle, E_{g,0(\alpha^2)}^{(2)}, E_{k(\alpha^2)}^{\pm(2)}$ correspond to the contribution of PDM of polar molecule, $|\Psi_{g,0(CRT)}^{(1)}\rangle, |\Psi_{k(CRT)}^{\pm(1)}\rangle, |\Psi_{g,0(CRT)}^{(2)}\rangle, |\Psi_{k(CRT)}^{\pm(2)}\rangle, E_{g,0(CRT)}^{(2)},$  $E_{k(CRT)}^{\pm(2)}$ correspond to the contribution of CRT, and $|\Psi_{g,0(\alpha,CRT)}^{(2)}\rangle, |\Psi_{k(\alpha,CRT)}^{\pm(2)}\rangle$ correspond to the combined contribution of both the PDM and the CRT. Where the first-order wavefunction correction $|\Psi_{g,0(\alpha)}^{(1)}\rangle, |\Psi_{k(\alpha)}^{\pm(1)}\rangle$ are proportional to $\alpha$, and the first-order energy correction $E_{g,0}^{(1)}, E_{k}^{\pm(1)}$ equals zero; while the second-order wavefunction and energy correction $|\Psi_{g,0(\alpha^2)}^{(2)}\rangle, |\Psi_{k(\alpha^2)}^{\pm(2)}\rangle, E_{g,0(\alpha^2)}^{(2)}, E_{k(\alpha^2)}^{\pm(2)}$ are proportional to $\alpha^2$, and $|\Psi_{g,0(\alpha,CRT)}^{(2)}\rangle, |\Psi_{k(\alpha,CRT)}^{\pm(2)}\rangle$ are proportional to $\alpha$. In particular, when $\alpha=0$, all of these results reduce to those of two-level atom Rabi model case, hence the DSP theory is also a new analytical method to treat atom Rabi model.

In the following, we will discuss the validity of our approach. The definite meaning of a small $H^{\prime}$ is that the matrix element of perturbation operator between the perturbing state $|\varphi_n^{\pm}\rangle$ and the unperturbed state $|\varphi_k^{\pm}\rangle$ should be much smaller than the energy difference between the two states: $\left|\frac{V_{nk}^{\pm\pm}}{\varepsilon_n^{\pm}-\varepsilon_k^{\pm}}\right|<<1$. Thus, the power series of $E$ and $|\Psi\rangle$ will convergence quickly, and we only need to calculate the first few terms to obtain quite accurate results. Through calculation, our DSP theory up to second-order correction is valid in a certain range of both normalized coupling strength $f=\lambda/\omega_c$ and $\alpha$ value: $f<<0.4, |\alpha|<<\frac{2}{(2+\sqrt3)f}-2$ in the resonant case, and the range of validity of $\alpha$ narrows as $f$ value increases. In addition, we focus on coherent coupling here in this DSP theory, and the involvement of decay of both cavity field and the molecule will be discussed in our future paper.

\section{The effect of PDM on the population and energy level}
In this section we will discuss the effect of PDM on the energy level and population of the system. The energy levels of the system are displayed in Fig. 1. The first column stands for the uncoupled stystem, while the second column stands for the dressed energy levels for the unperturbed system under the RWA, and the third column stands for the dressed energy levels for the perturbed system beyond the RWA. Suppose the system is in one of the eigenstates before being perturbed, for example in the ground state $|g,0\rangle$, under the framework of the RWA, the ground state of the system is always uncoupled from other states and thereby remains unchanged. Yet when the CRTs and PDM are involved in the treatment, the ground state will always overlap with the upper level and the photon in the ground state is no longer in a vacuum state, but now contains molecule-cavity coupling \cite{fb}. Under first-order perturbation, PDM lead the coupling between the states $|g,0\rangle$ and $|\varphi_0^{\pm}\rangle$, while CRT lead the coupling between state $|g,0\rangle$ and $|\varphi_1^{\pm}\rangle$ according to appendix (B.1-B.2). Furthermore, under second-order perturbation, PDM induce the transition between the states $|\varphi_0^{\pm}\rangle$ and $|\varphi_1^{\pm}\rangle$, $|\varphi_1^{\pm}\rangle$ and $|\varphi_2^{\pm}\rangle$, while CRT induce the transition between the states $|\varphi_0^{\pm}\rangle$ and $|\varphi_2^{\pm}\rangle$, $|\varphi_1^{\pm}\rangle$ and $|\varphi_3^{\pm}\rangle$ according to appendix (B.3-B.5). With the increase of the order of the perturbation, there will be the coupling between higher order dressed states.
\begin{figure}
\includegraphics[angle=0,width=12.0cm]{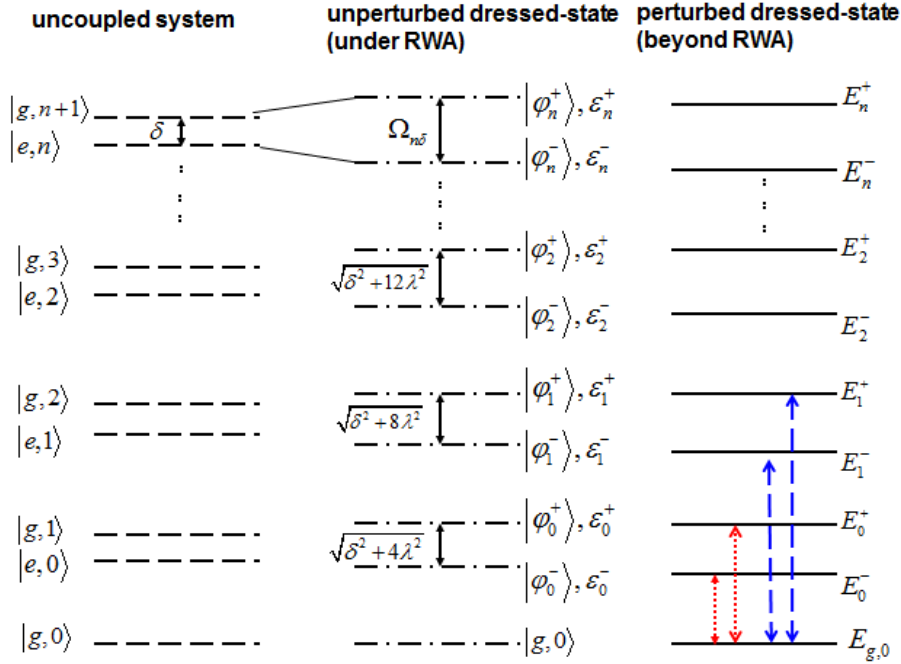}
\caption{(Color online) The dressed energy level of the system. The black solid arrows correspond to the coupling produced by the rotating wave terms, the blue dashed arrows correspond to the coupling produced by CRT and the red dotted arrows correspond to the coupling produced by PDM.}
\end{figure}

\begin{figure}
\includegraphics[width=0.35\columnwidth]{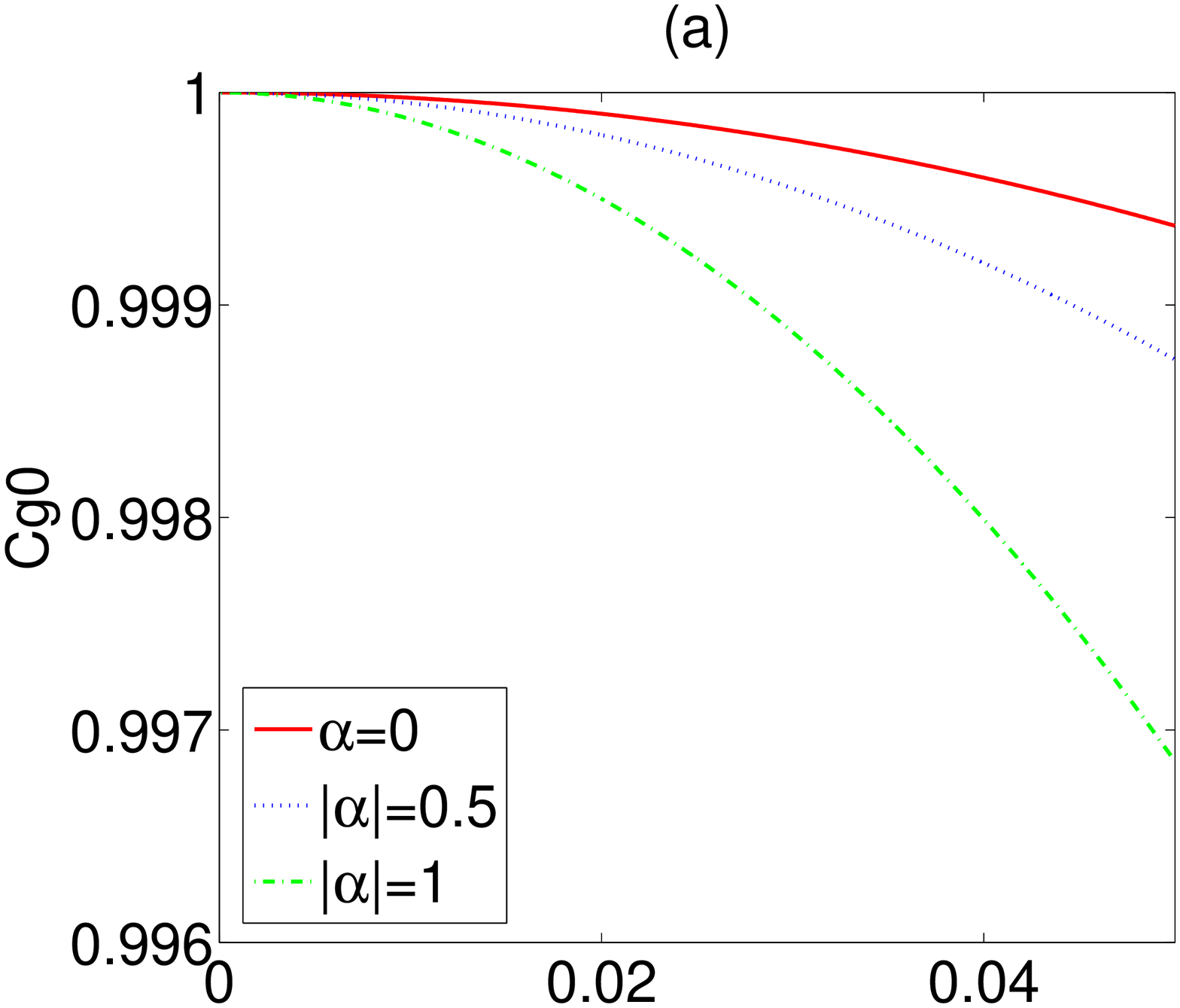}
\includegraphics[width=0.35\columnwidth]{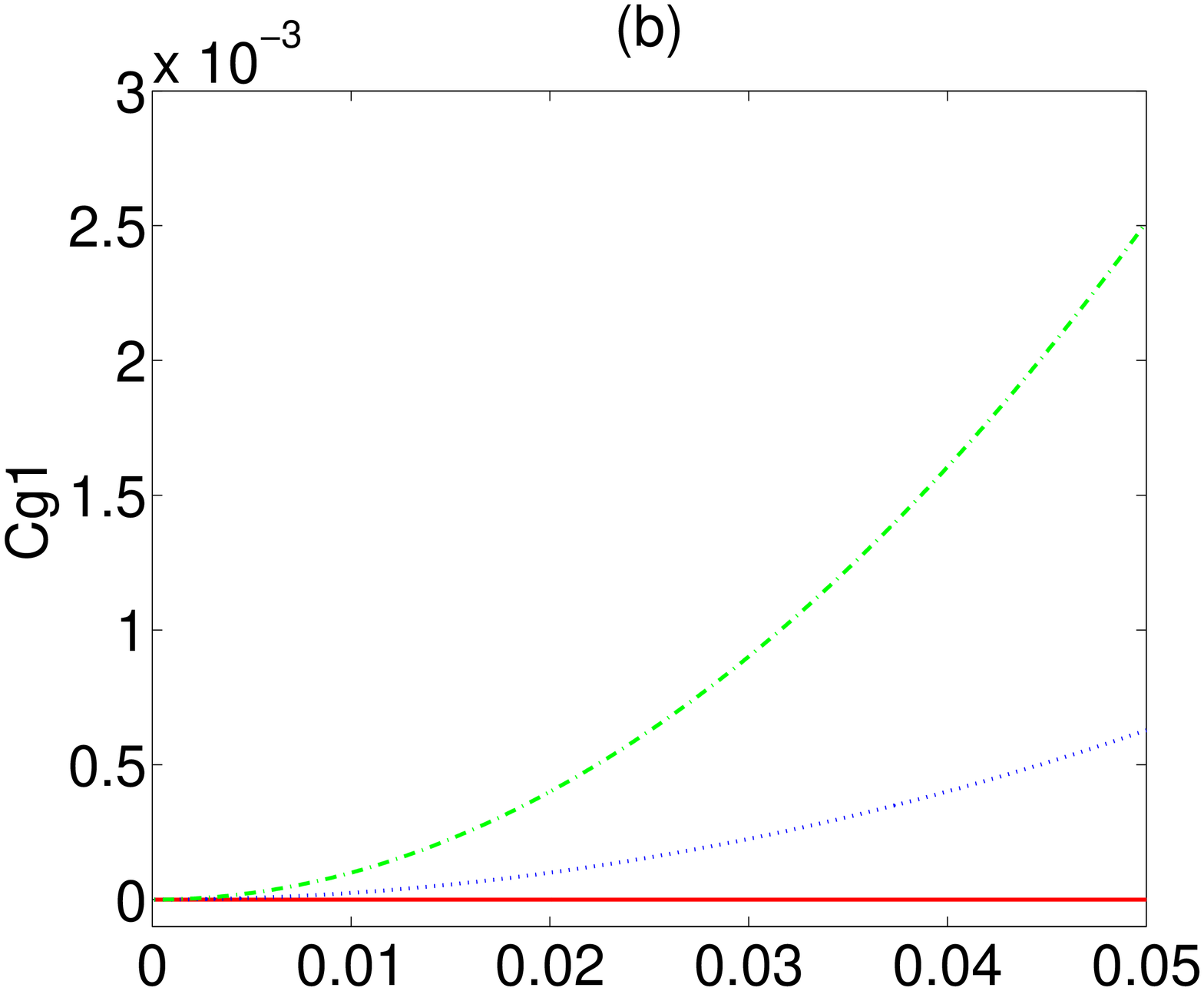}
\includegraphics[width=0.35\columnwidth]{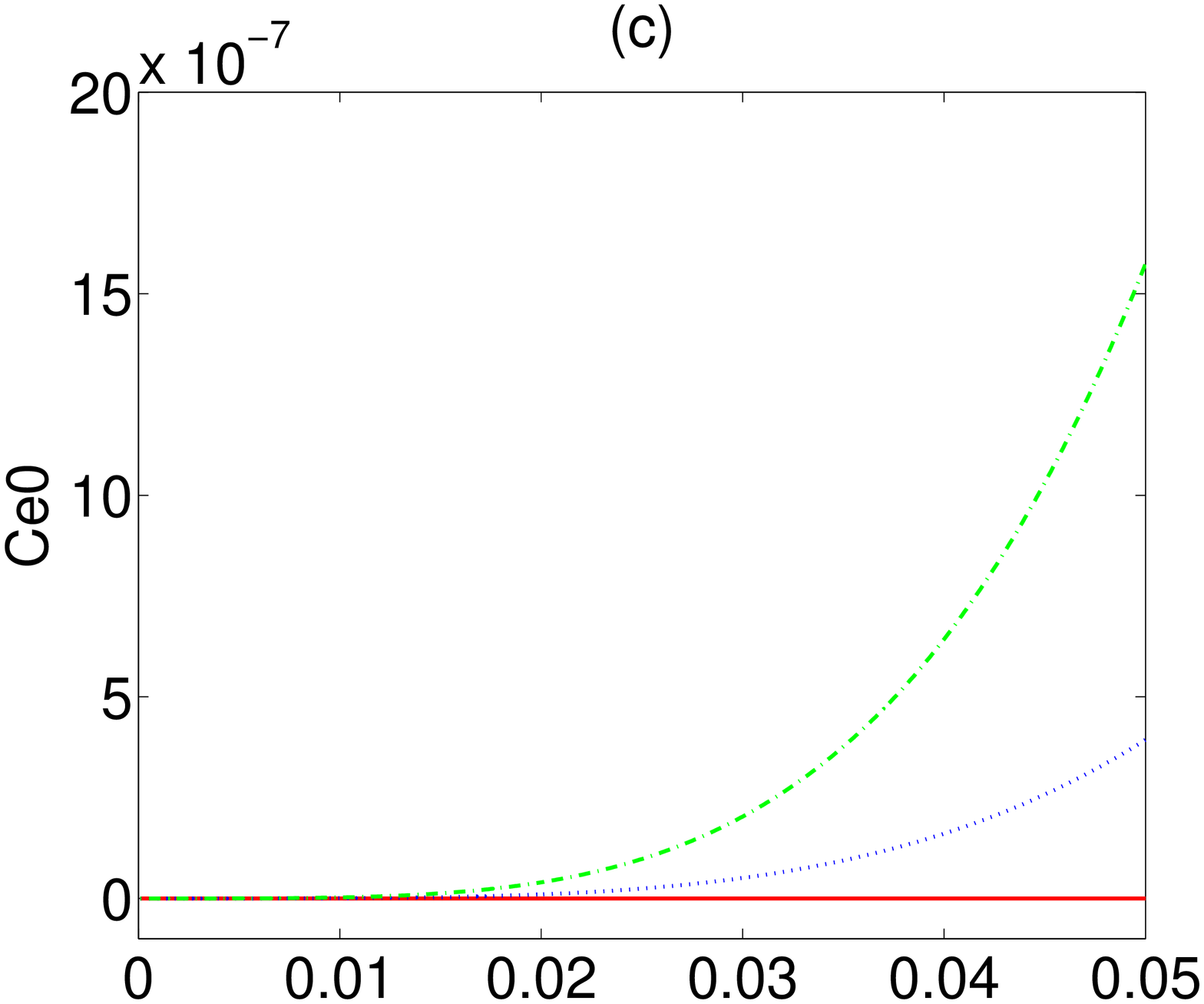}
\includegraphics[width=0.35\columnwidth]{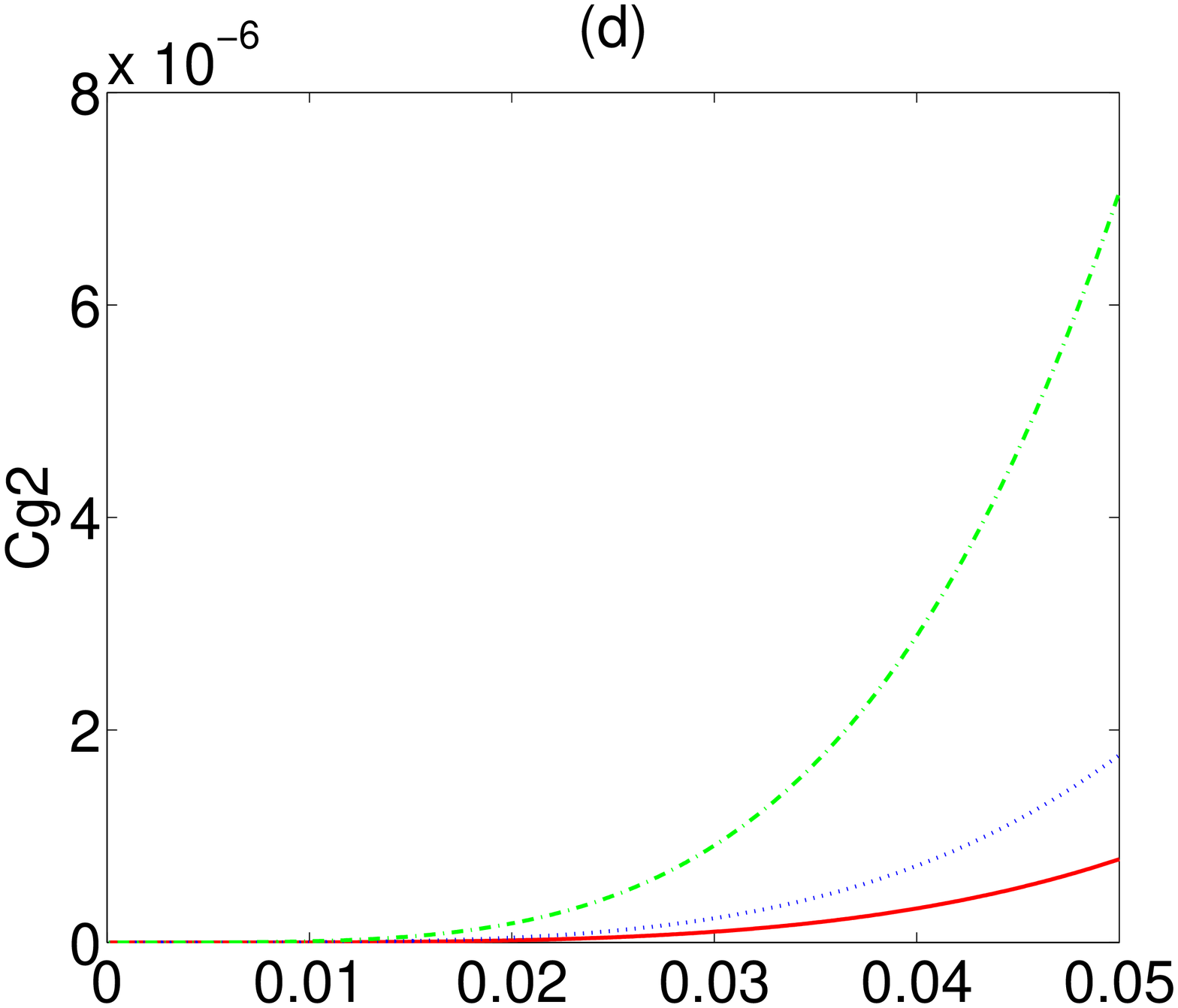}
\includegraphics[width=0.35\columnwidth]{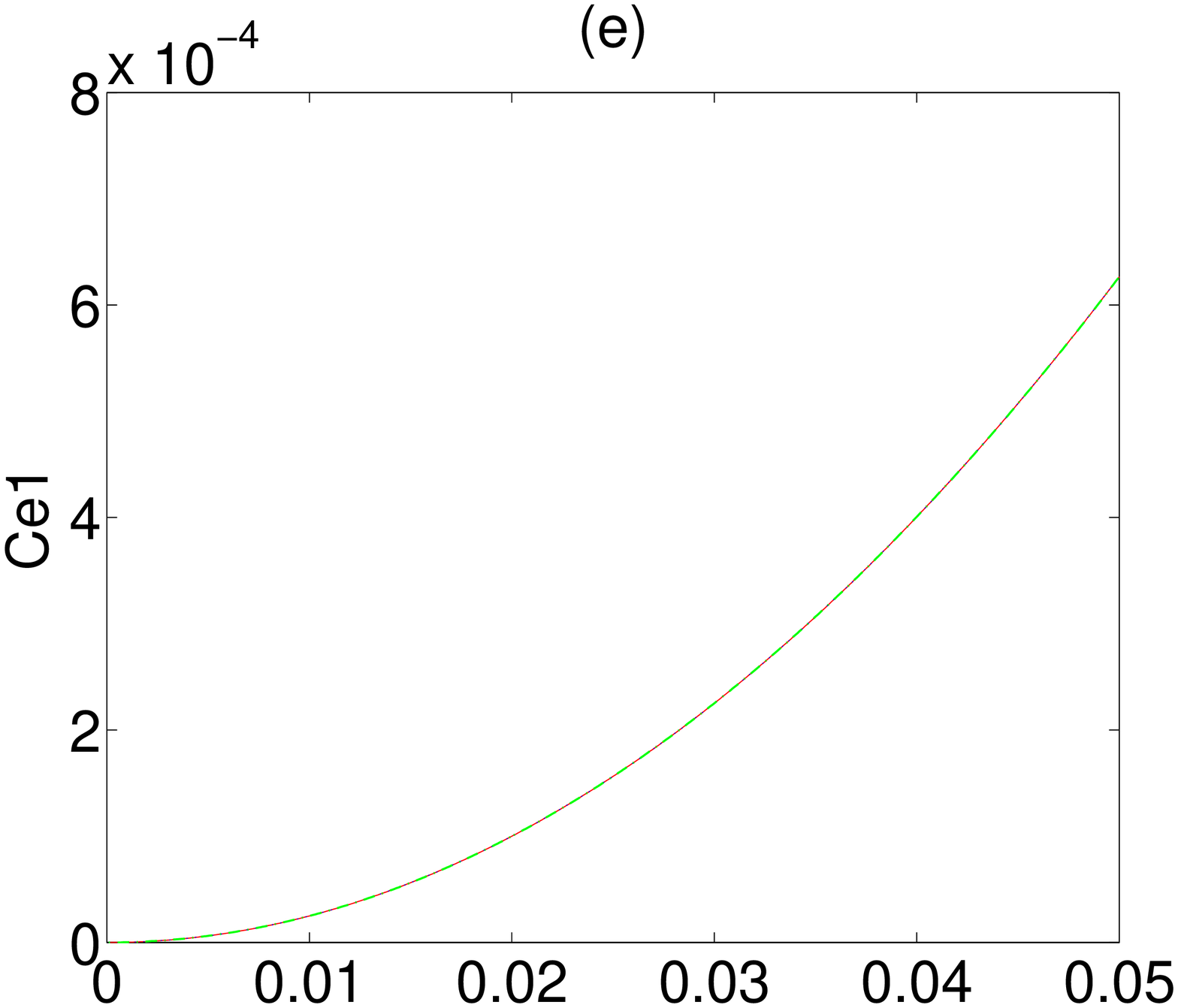}
\includegraphics[width=0.35\columnwidth]{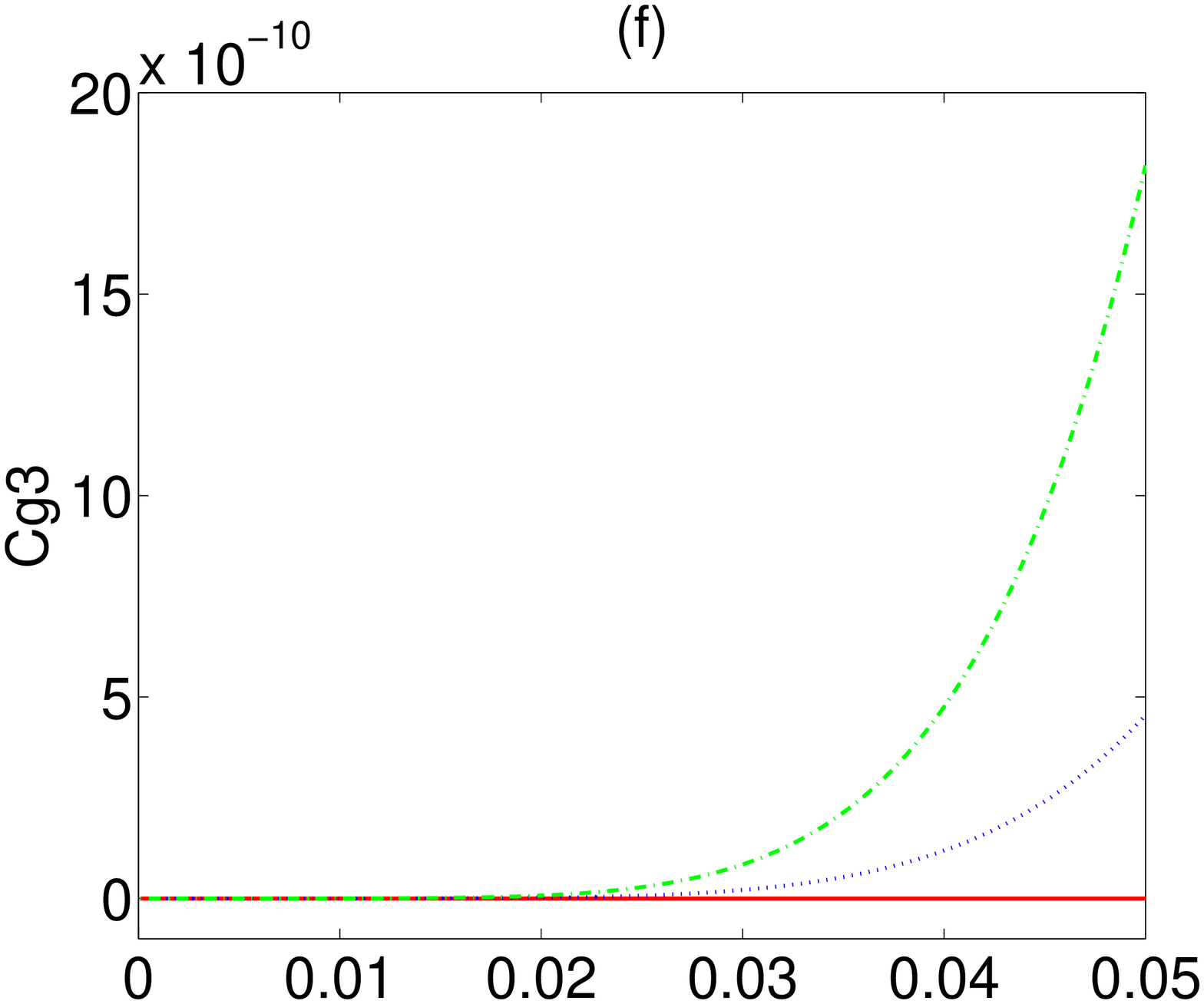}
\includegraphics[width=0.35\columnwidth]{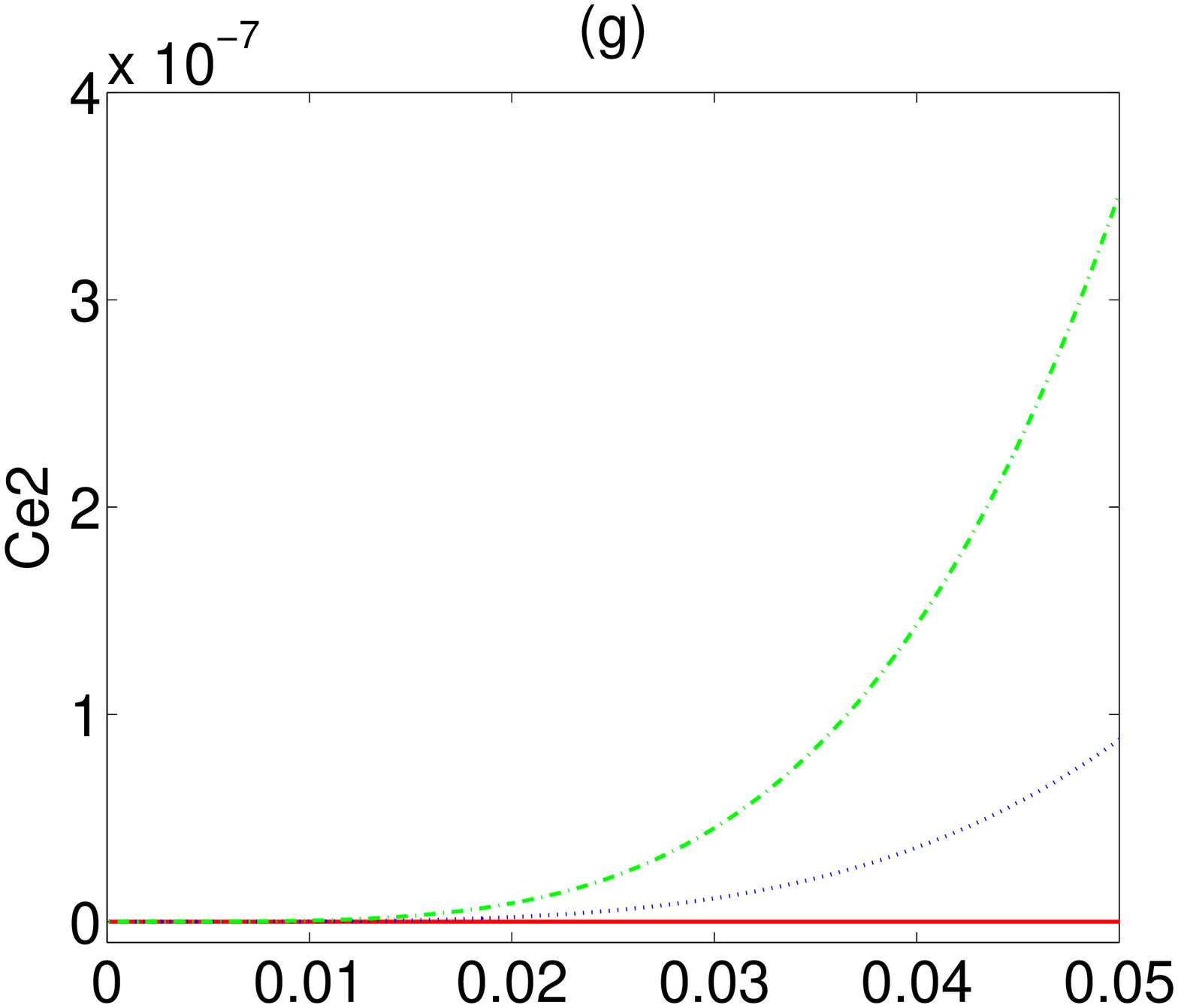}
\includegraphics[width=0.35\columnwidth]{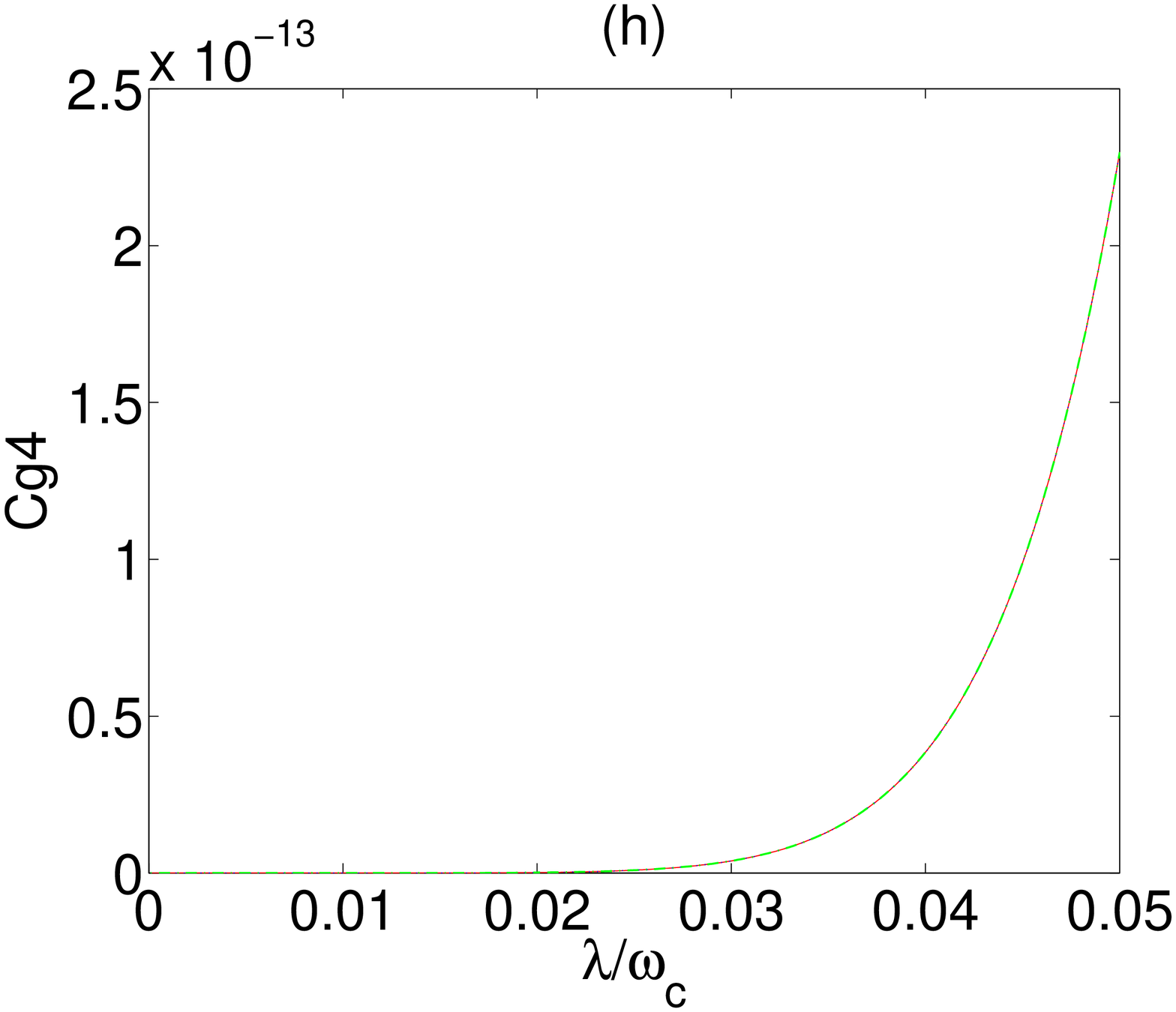}
\includegraphics[width=0.35\columnwidth]{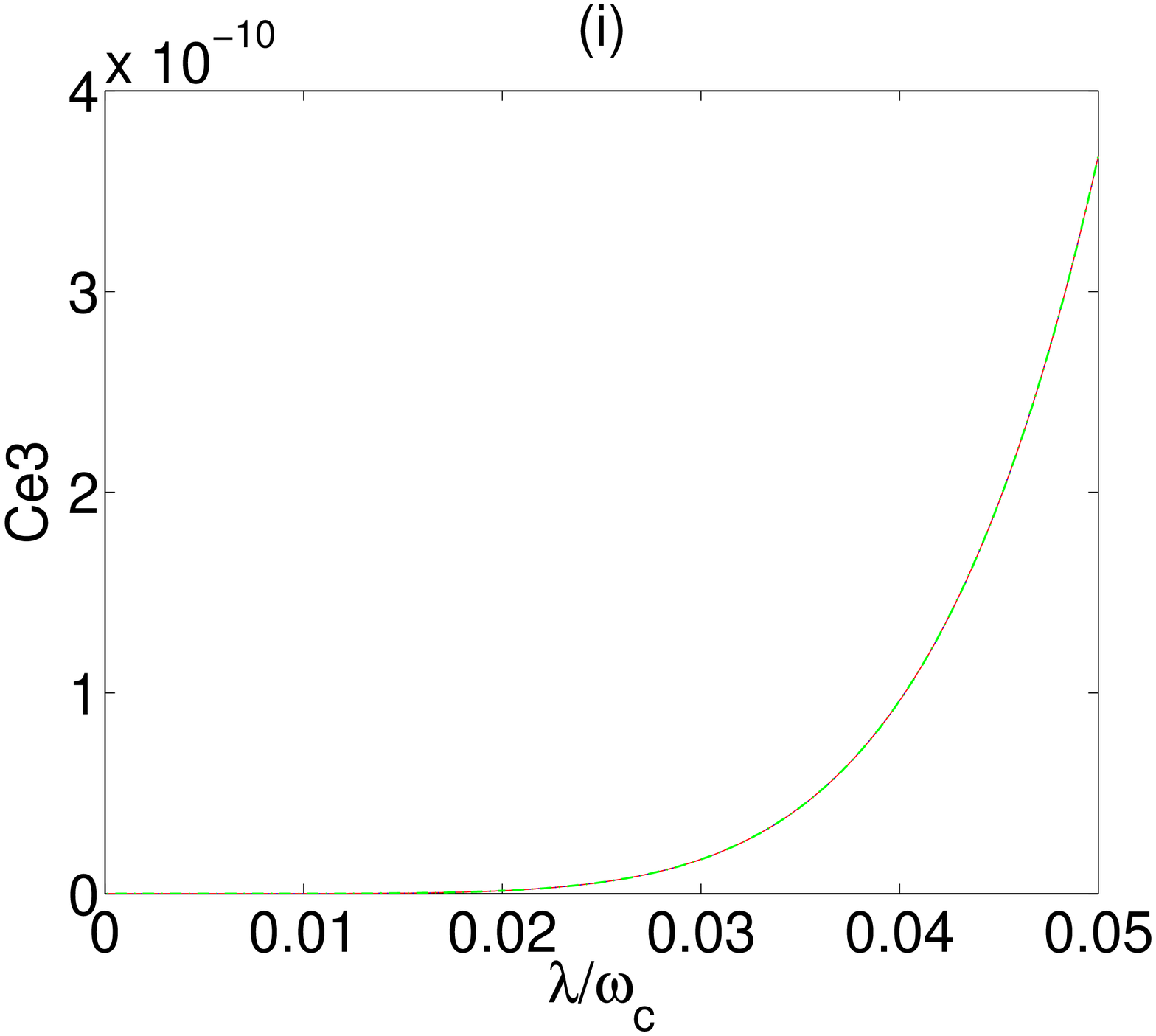}
\caption{(Color online) The population probability of different energy level $(a)|g,0\rangle$, $(b)|g,1\rangle$, $(c)|e,0\rangle$, $(d)|g,2\rangle$, $(e)|e,1\rangle$, $(f)|g,3\rangle$, $(g)|e,2\rangle$, $(h)|g,4\rangle$, $(i)|e,3\rangle$ against normalized coupling constant $f=\lambda/\omega_c$ for different $\alpha$ in the resonant case. Red solid lines represent $\alpha=0$, blue dotted lines represent $|\alpha|=0.5$, and green dot-dashed lines represent $|\alpha|=1$. In (e), (h) and (i), the differences among the curves are very small.}
\end{figure}

In order to make the transition relation between the energy levels more explicit, we can furthermore expand the dressed energy level $|\varphi_n^{\pm}\rangle$ in terms of $\{|e,n\rangle,|g,n+1\rangle\}$. The population probability of ground state $|g,0\rangle$ and higher energy levels $|e,n\rangle,|g,n+1\rangle$ (n=0,1,2,3) against normalized coupling constant $f=\lambda/\omega_c$ for different $\alpha$ in the resonant case are exhibited in Fig. 2. In the whole paper, we set $\omega_c=1$ for simplicity. As shown in the Fig. 2(a), when the coupling strength becomes greater, the population probability of ground state $|g,0\rangle$ will decrease, and the higher order states' population probability will increase, yet the populations of the higher order states are much smaller than that of the lower order one according to Fig. 2(b)-(i). Besides, when $\alpha=0$, CRT only lead the coupling between state $|g,0\rangle$ and $|\varphi_1^{\pm}\rangle$, $|\varphi_1^{\pm}\rangle$ and $|\varphi_3^{\pm}\rangle$ as shown in  Fig. 2(a),(d),(e),(h) and (i), while PDM furthermore lead the coupling between the states $|g,0\rangle$ and $|\varphi_0^{\pm}\rangle$, $|\varphi_0^{\pm}\rangle$ and $|\varphi_1^{\pm}\rangle$, $|\varphi_1^{\pm}\rangle$ and $|\varphi_2^{\pm}\rangle$. As $|\alpha|$ increases, the coupling between dressed states will strengthen.

Below we will present a detailed discussion on the energy sprectrum of the polar molecule cavity QED system for various values of $f$ and $\alpha$ to further demonstrate the validity of our analytical approach. Figure 3 shows the seven lower-lying energy levels $E_{g,0}, E_k^{\pm},k=(0,1,2)$ as a function of $f$, $\alpha$ and detuning $\delta$, respectively. We compare eigenvalues obtained by our DSP theory with those by numerical diagonalization in a truncated finite-dimensional Hilbert space \cite{qh}. It is shown that for the parameter regimes with $f<0.05$, $|\alpha|<1$ and a wide range of detuning values $-1<\delta/\omega_c<1$, our analytical energy spectrums are in good agreement with the direct numerical simulation. Therefore, the DSP theory provides an accurate analytical expression to the energy spectrum of the two-level polar molecule cavity QED system in a certain range of both $f$ and $\alpha$ value. Note that in the present experimentally accessible systems the maximum value for the coupling strength is generally realized in the superconducting circuit resonator \cite{23pf} and organic microcavity \cite{jrt}, which is  around $f\sim0.1$. So it should be practically interesting to find a good solution in this coupling regime. As for the case of ultrastrong coupling regime, we will discuss it in our future work. In Fig. 3(a), when $f=0$, the ground state is nondegenerate and each higher energy level is doubly degenerate. The separation between the levels is $\omega_c$, which is also equal to $\omega_0$ in the resonant case. As $f$ increases, the energies of the ground state $E_{g,0}$ and $E_k^{-}$ gradually decrease, while the energies of $E_k^{+}$ gradually increase within the valid regime $f<0.05$. According to Fig. 3(b), in the case of coupling $f=0.05$, there are small variations of the energies within the valid regime $|\alpha|<1$. In Fig. 3(c), the energy spectrum exhibits avoided crossings adjacent to $\delta=0$. The shift of the resonance frequency is so-called B-S shift \cite{jh}, which could also be described by the energy shift of the level transition with the consideration of the CRT in the strong coupling regime \cite{23pf,ly}. For instance, in Ref. \cite{jh}, they noted that CRTs yield second-order corrections in coupling strength to the resonance frequency. Also, in Ref. \cite{th}, it is shown that the terms higher than second order of laser Rabi frequency in the B-S shift cannot be neglected for giant-dipole (large permanent dipole difference) molecules under an intense oscillating field. We will discuss the magnitude of it in detail in the following passage.

\begin{figure}
\includegraphics[width=0.55\columnwidth]{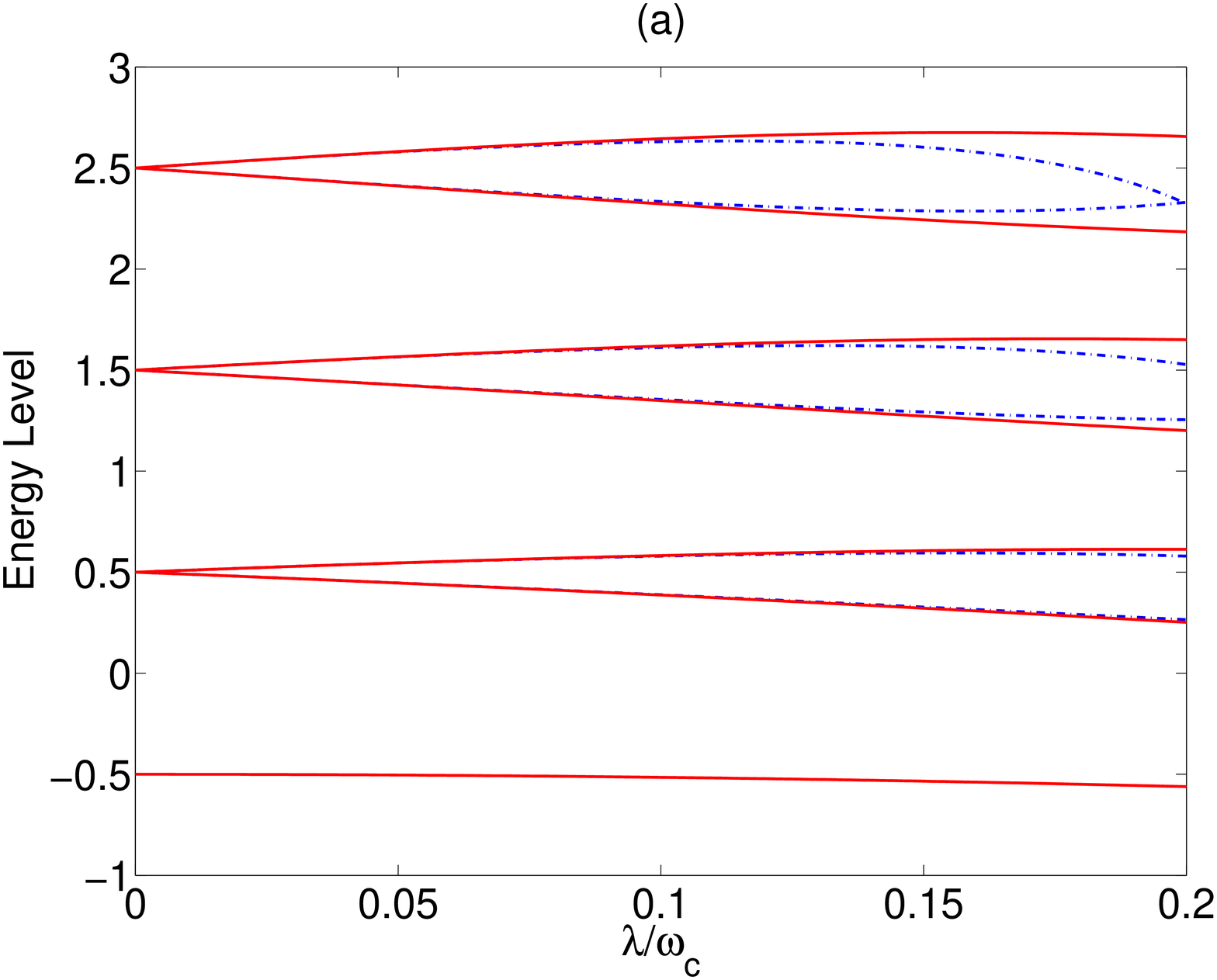}
\includegraphics[width=0.55\columnwidth]{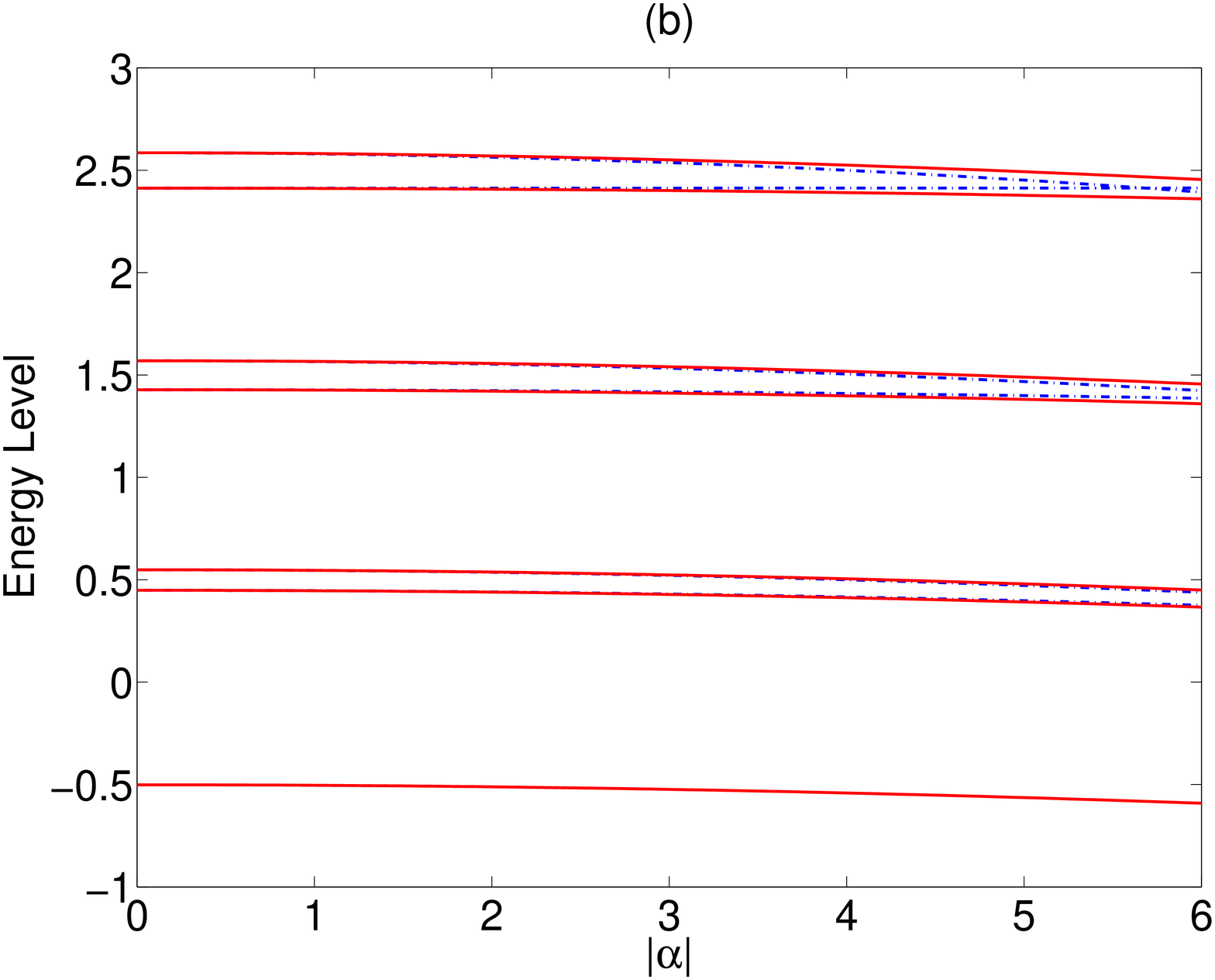}
\includegraphics[width=0.55\columnwidth]{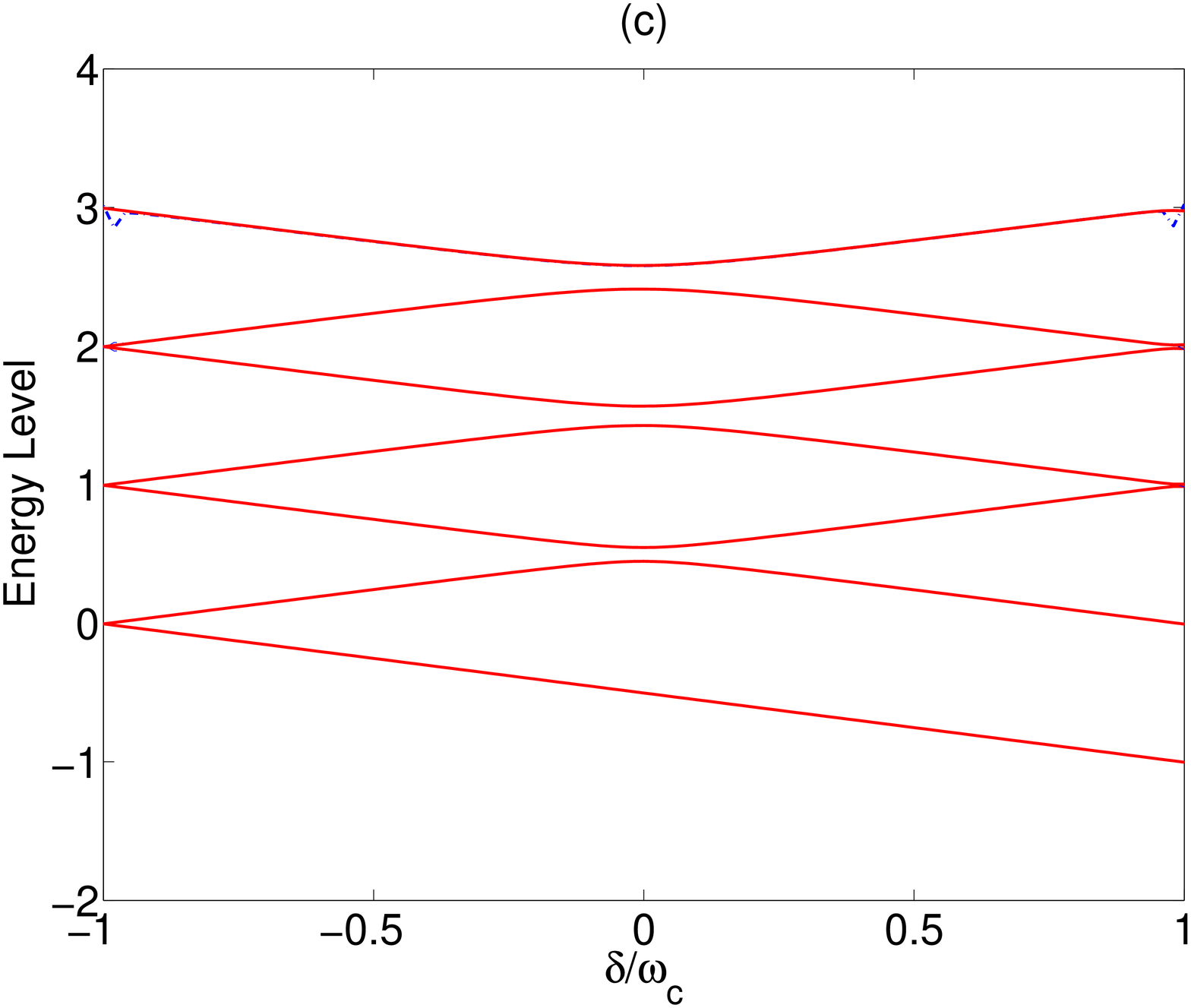}
\caption{(Color online) The seven lower-lying energy levels $E_{g,0}, E_k^{\pm},k=(0,1,2)$ as a function of $f$, $\alpha$ and detuning $\delta$, respectively. (a) $\delta=0, |\alpha|=1$; (b) $\delta=0, f=0.05$; (c) $|\alpha|=1, f=0.05$. Blue dot-dashed lines represent the results of our dressed perturbation theory, while red solid lines represent the results of numerical simulation. For the parameter regimes with $f<0.05$, $|\alpha|<1$ and a wide range of detuning values $-1<\delta/\omega_c<1$, our analytical energy spectrums are in good agreement with the direct numerical simulation.}
\end{figure}

From the analytical expressions of the energy level Eqs. (10)-(15), we can also find that both the CRT and the PDM result in the energy level shift from the unperturbed energy, and the magnitude of the energy level shift is dependent on coupling strength $\lambda$, normalized permanent dipole difference $\alpha$, cavity field frequency $\omega_c$, molecule transition frequency $\omega_0$ and photon number $k$, where the energy shift $E_{g,0(CRT)}^{(2)}, E_{k(CRT)}^{\pm(2)}$ induced by the CRT is approximately proportional to the $\lambda^2$, and the energy shift $E_{g,0(\alpha^2)}^{(2)}, E_{k(\alpha^2)}^{\pm(2)}$ induced by the PDM is proportional to $\alpha^2$.

\begin{figure}
\includegraphics[width=0.55\columnwidth]{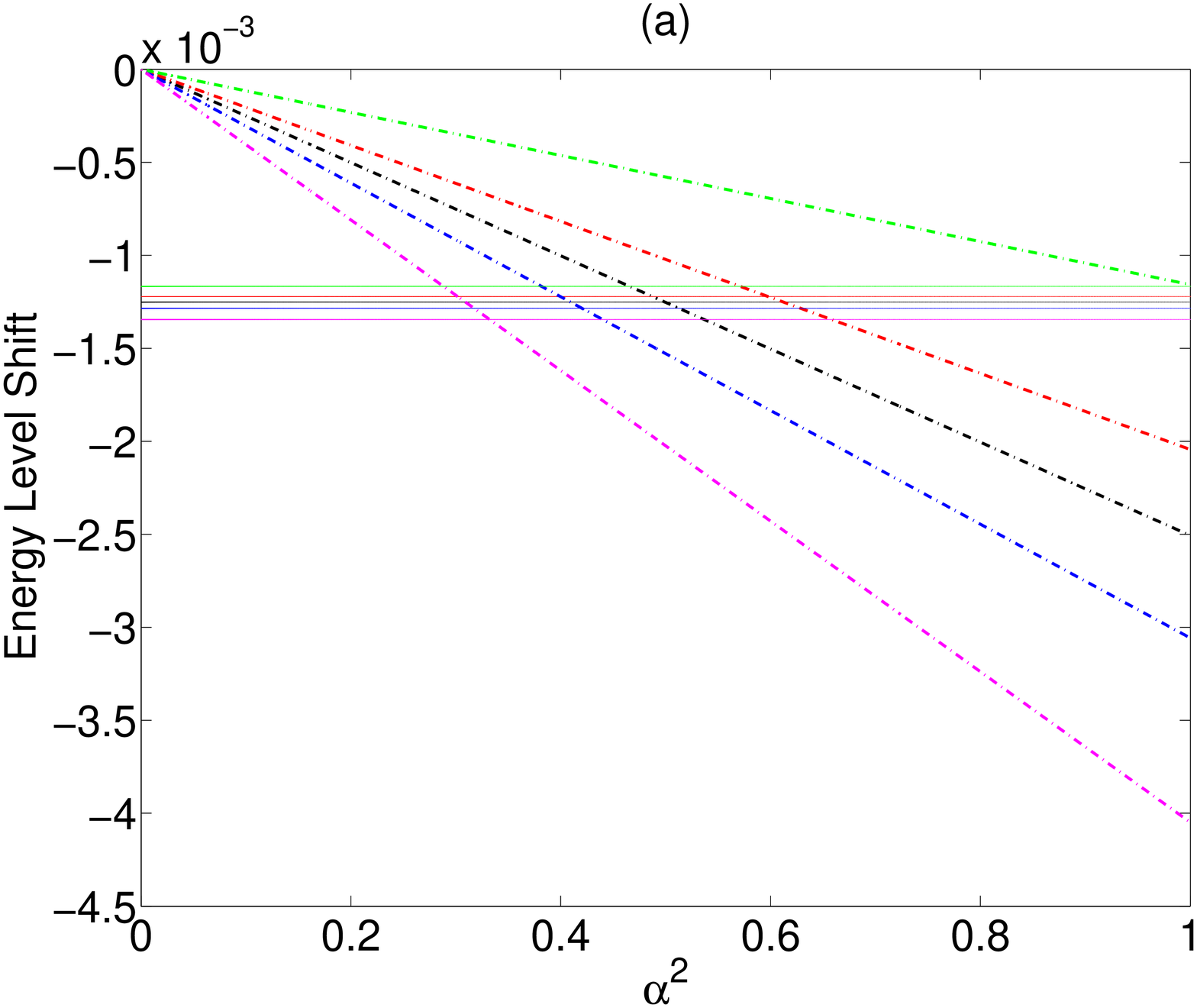}
\includegraphics[width=0.55\columnwidth]{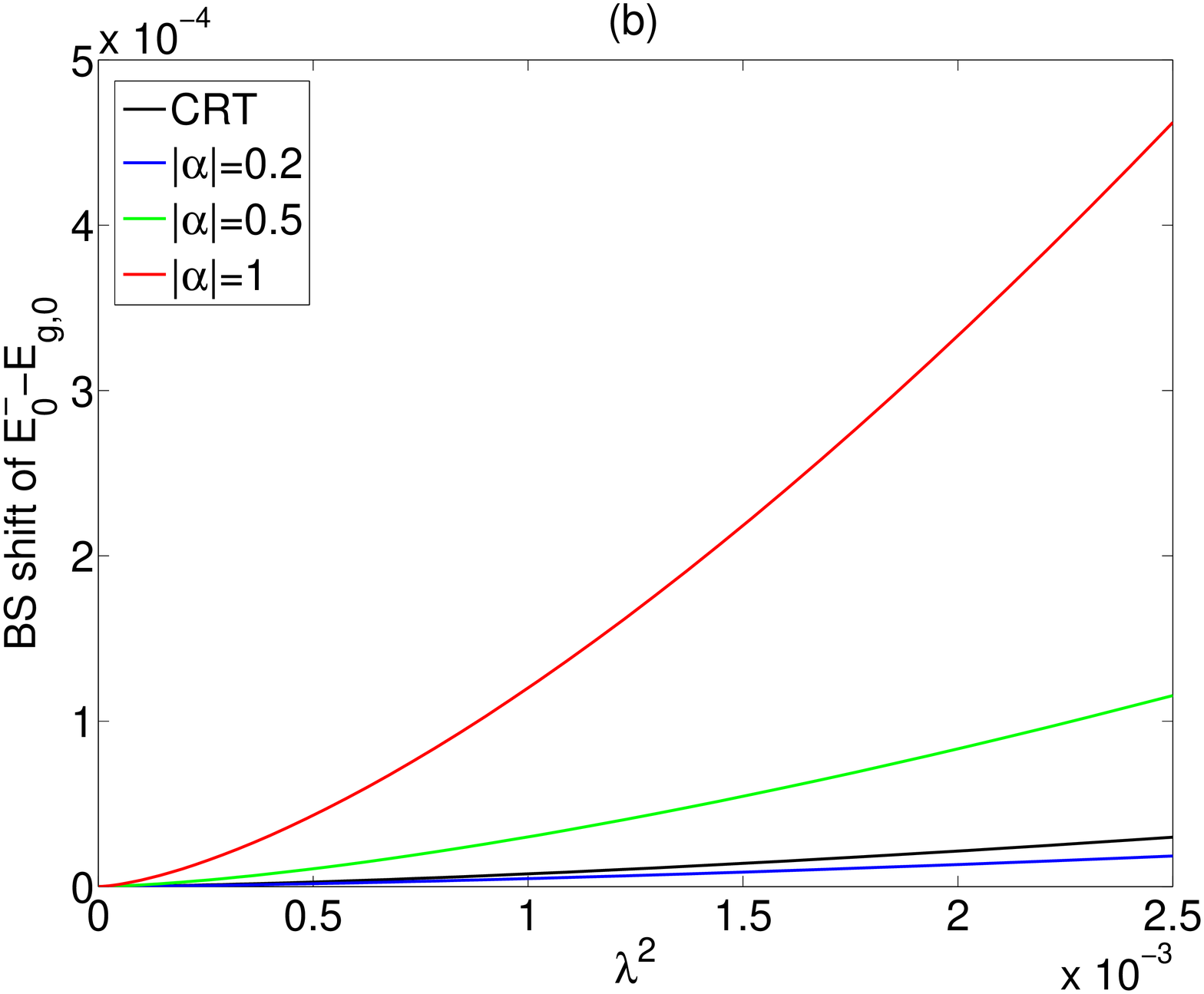}
\includegraphics[width=0.55\columnwidth]{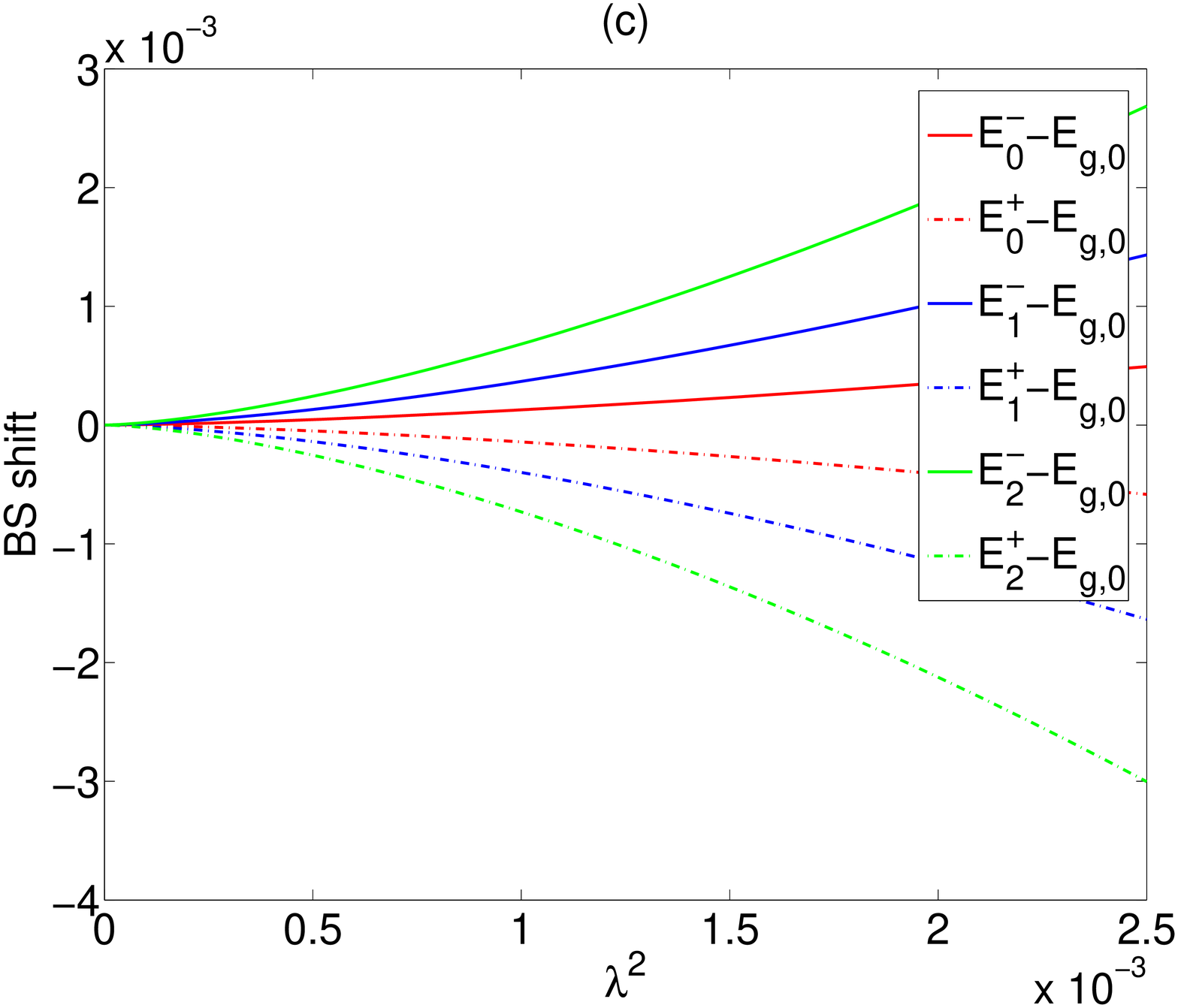}
\caption{(Color online)(a) The respective contributions of the PDM and the CRT in the energy shift of the five lower-lying energy levels as a function of $\alpha^2$ for $f=0.05$ in the resonant case, where those curves parallel to abscissa axis represent energy level shift caused by the CRT. The black lines: $E_{g,0(\alpha^2)}^{(2)}$ and $E_{g,0(CRT)}^{(2)}$, the red lines: $E_{0(\alpha^2)}^{-(2)}$ and $E_{0(CRT)}^{-(2)}$, the blue lines: $E_{0(\alpha^2)}^{+(2)}$ and $E_{0(CRT)}^{+(2)}$, the green lines: $E_{1(\alpha^2)}^{-(2)}$ and $E_{1(CRT)}^{-(2)}$, the magenta lines: $E_{1(\alpha^2)}^{+(2)}$ and $E_{1(CRT)}^{+(2)}$. (b) The B-S shift of the transition $E_0^{-}\rightarrow E_{g,0}$ for the different $\alpha$ as a function of $\lambda^2$ with detuning $\delta=0$. (c) The total B-S shift caused by PDM and CRT with the transition $E_k^{\pm}\rightarrow E_{g,0}$ ($k=0,1,2$) for $|\alpha|=1$ as a function of $\lambda^2$ with detuning $\delta=0$.}
\end{figure}

In Figure 4(a), we plot the respective contributions of the PDM and the CRT in the energy shift of the five lower-lying energy levels $E_g, E_k^{\pm},k=(0,1)$ as a function of $\alpha^2$ for $f=0.05$ in the case of exact resonance $\delta=0$. For polar molecule with relatively small $|\alpha|$, the contribution of PDM in the energy shift is less than that of the CRT. While for some aromatic molecules frequently used in lasers and nonlinear optics \cite{27rb} and organic compounds presenting significant nonlinear optical properties \cite{og}, the PDM difference between the electronic excited state and ground state could be very large, which is so-called giant dipole molecule, $|\alpha|$ value could well exceed 1. Then from Fig. 4(a), the contribution of PDM in the energy shift could surpass that of the CRT. This conclusion can be also drawn from Fig. 4(b). The B-S shift can be also obtained by means of Eqs. (10)-(15). In Fig. 4(b), the smallest B-S shift of the transition $E_0^{-}\rightarrow E_{g,0}$ for the different $\alpha$ as a function of $\lambda^2$ with detuning $\delta=0$ is plotted. It is shown that the B-S shift caused by CRT is approximately proportional to $\lambda^2$, and those caused by PDM are dependent on the higher terms of $\lambda$. For giant dipole molecule, the contribution of PDM could produce greater BS shift than that of the CRT, especially with the increasing coupling strength $\lambda$. The total B-S shift caused by PDM and CRT with the transition $E_k^{\pm}\rightarrow E_{g,0}$ ($k=0,1,2$) for $|\alpha|=1$ as a function of $\lambda^2$ with detuning $\delta=0$ is shown in Fig. 4(c). For the transition between the higher excited state and the ground state $E_k^{\pm}\rightarrow E_{g,0}$, the magnitude of the BS shift becomes larger as k increases, and its sign changes alternatively with $\mp$.

The parameters used for calculation here are within the range of practical molecules. For instance, in the coupling of electronic transitions of SrF \cite{7ara} and diphenyl molecules $[N(CH_3)_2-C_6H_4-(C\equiv C)_2-C_6H_4-NO_2]$ \cite{shnc} with visible field, we calculate the BS shift of the transition $E_0^{-}\rightarrow E_{g,0}$ caused by the PDM and CRT, respectively for $f=0.05$ in the resonant case, and the results are given in table I. With a greater $\alpha$, the contribution of PDM could produce a greater BS shift than that caused by CRT.

{\tiny
\begin{table}[h]
\centering
\caption{The BS shift $(cm^{-1})$ caused by the PDM and CRT in the coupling of electronic transitions of polar molecules with visible field. $f=\lambda/\omega_c=0.05$ in the resonant case $\omega_c=\omega_0$.}
\begin{tabular}{|l|l|l|l|l|l|}
\hline
molecule & transition ($\mu_{gg},\mu_{ee},\mu_{ge}$ (D)) & $\alpha$ & $\omega_0(cm^{-1})$ &  PDM shift & CRT shift \\\hline
SrF      & $X^2\Sigma-A^2\Pi$ (3.482, 2.059, 6.227) & -0.114 & $1.52\times10^4$ & 0.091 & 0.454 \\\hline
diphenyl & (1-2) (10.67, 16.66, 9.13) & 0.328 & $3.05\times10^4$ & 1.53 & 0.91 \\\hline
         & (1-4) (10.67, 12.99, 3.9) & 0.297 & $3.85\times10^4$ & 1.57 & 1.151  \\\hline
         & (1-7) (10.67, 14.07, 3.37) & 0.504 & $4.89\times10^4$ & 5.65 & 1.461 \\\hline
\end{tabular}
\end{table}}

\section{CONCLUSIONS}
In the present paper, we apply the dressed-state perturbation theory beyond the RWA to investigate the interaction between a two-level polar molecule and a quantized cavity field. Analytical expressions have been derived for both the ground state and the excited state energy spectrums and wave functions of the system, where the contribution of permanent dipole moments and the counter-rotating wave term can be shown separately. The validity of our approach is discussed by comparing with the numerical diagonolization method. For the parameter regimes with $f<0.05$, $|\alpha|<1$ and a wide range of detuning values $-1<\delta<1$, our analytical results agree well with the direct numerical simulation. It implies that our presented DSP theory is valid for the current experimental conditions with strong coupling. From the expression of the wave function, we could clearly obtain the population of the different dressed-state energy level and the transition relation between the energy levels. The respective effect of PDM and CRT on both the population and energy level of the system are analyzed. Comparing with CRT coupling, PDM induce the coupling of more dressed states, and as $|\alpha|$ increases, the coupling between dressed states will strengthen. Furthermore, both the CRT and the PDM result in the energy shift of the dressed states from those of the JC model Hamiltonian, and the energy shift caused by the CRT is approximately proportional to the square of coupling strength, while the energy shift caused by the PDM is proportional to the square of normalized permanent dipole difference and dependent on the higher terms of coupling strength. For giant dipole molecule cavity QED, it is expected that the contribution of permanent dipole moment could produce a greater Bloch-Siegert shift than that of the counter-rotating wave term. Moreover, our method could also be extended to the solution of two-level atom Rabi model Hamiltonian beyond the RWA by setting $\alpha=0$.

In addition, much theoretical and experimental attention has been devoted to the effect of the counter-rotating wave terms, which plays an important role in the strong-coupling regime and result in some novel quantum mechanical phenomena, for instance, the modification of dynamics of population inversion \cite{28jl}, preparing non-classical state \cite{29sa}, quantum irreversibility and chaos \cite{30lb}, quantum phase transition \cite{31rf}, the simulation of dynamical Casimir effect in semiconductor microcavity or superconducting circuit QED \cite{32vv}, and the production of photon from vacuum \cite{33tw} etc. Comparing the contribution of permanent dipole moment with that of the counter-rotating wave terms will lead to beneficial discussion, and we will work on it in the future paper.

\ack{This work is supported by the Strategic Priority Research Program of the Chinese Academy of Sciences, Grant No. XDB01010200, the Hundred Talents Program of the Chinese Academy of Sciences, Grant No. Y321311401, the National Natural Sciences Foundation of China (Grant No. 61475139, No. 11347147 and No. 11247014), 973 program of China (Grant No. 2013CB329501), and the Zhejiang Provincial Natural Science Foundation under Grant No. LQ13A040006.}

\begin{appendix}
\section{dressed-state perturbation theory beyond the RWA}
In order to calculate the various order approximation, we introduce a real parameter $j$ which varies continuously from zero to 1, and rewrite $H^{\prime}$ as $jV$:
\begin{equation}
H^{\prime}=jV=-j\lambda[\alpha\sigma_z(a^{\dag}+a)+a^{\dag}\sigma_{+}+a\sigma_{-}]
\end{equation}
Now to solve the whole quantum behavior of the system is to deal with new stationary problem:
\begin{eqnarray}
&& H|\Psi\rangle=E|\Psi\rangle,\\
&& \langle\Psi|\Psi\rangle=1,
\end{eqnarray}
The equation is hard to exactly solved, but we can obtain the correction on the base of the solution of the unperturbed system.
Inserting Eq.(1) to Eq.($A2$), we have:
\begin{eqnarray}
&&(E-\varepsilon_m^{+})a_m=
j\left[\sum_{n=0}^{\infty}(a_nV_{mn}^{++}+b_nV_{mn}^{+-})+c V_{mg}^{+0}\right]\label{} \\
&&(E-\varepsilon_m^{-})b_m=
j\left[\sum_{n=0}^{\infty}(a_nV_{mn}^{-+}+b_nV_{mn}^{--})+c V_{mg}^{-0}\right]\label{} \\
&&(E+\frac{\omega_0}{2})c=
j\left[\sum_{n=0}^{\infty}(a_nV_{gn}^{0+}+b_nV_{gn}^{0-})+c V_{gg}^{00}\right]\label{}
\end{eqnarray}
where $V_{mn}^{\pm\pm}=\langle\varphi_m^{\pm}|V|\varphi_n^{\pm}\rangle$ is the perturbed matrix element.

Both the probability amplitude $a_n,b_n,c$ and energy E are all related to the perturbation, dependent on the parameter j representing the strength of the perturbation. Expanding $a_n,b_n,c$ and energy E by the power series of j:
\begin{eqnarray}
&& a_n=a_n^{(0)}+ja_n^{(1)}+j^2 a_n^{(2)}+\cdots
=\sum_{r=0}^{\infty}j^r a_n^{(r)}\label{equationa} \\
&& b_n=b_n^{(0)}+jb_n^{(1)}+j^2 b_n^{(2)}+\cdots
=\sum_{r=0}^{\infty}j^r b_n^{(r)}\label{equationb}\\
&& c=c^{(0)}+jc^{(1)}+j^2 c^{(2)}+\cdots
=\sum_{r=0}^{\infty}j^r c^{(r)}\label{equationb}\\
&& E=E^{(0)}+jE^{(1)}+j^2 E^{(2)}+\cdots
=\sum_{r=0}^{\infty}j^r E^{(r)}\label{equationc}
\end{eqnarray}

Inserting Eqs.$(A5a-d)$ into Eqs.$(A4a-c)$, we can have series equations that various-order approximated wave function $a_n^{(r)}, b_n^{(r)}$, $c^{(r)}$  and energy $E^{(r)}$ should satisfy.

\section{the wave function of ground state}
The first-order wave function correction is given by:$|\Psi_{g,0}^{(1)}\rangle=|\Psi_{g,0(\alpha)}^{(1)}\rangle+|\Psi_{g,0(CRT)}^{(1)}\rangle$, and the second-order wave function correction is: $|\Psi_{g,0}^{(2)}\rangle=|\Psi_{g,0(\alpha^2)}^{(2)}\rangle+|\Psi_{g,0(CRT)}^{(2)}\rangle+|\Psi_{g,0(\alpha,CRT)}^{(2)}\rangle$, where
\begin{eqnarray}
|\Psi_{g,0(\alpha)}^{(1)}\rangle= -\lambda\alpha\left(\frac{cos\theta_0}{\frac{\omega_0}{2}+\varepsilon_0^{+}}|\varphi_0^{+}\rangle-\frac{sin\theta_0}{\frac{\omega_0}{2}+\varepsilon_0^{-}}|\varphi_0^{-}\rangle\right)
\end{eqnarray}

\begin{eqnarray}
|\Psi_{g,0(CRT)}^{(1)}\rangle= \lambda\left(\frac{sin\theta_1}{\frac{\omega_0}{2}+\varepsilon_1^{+}}|\varphi_1^{+}\rangle+\frac{cos\theta_1}{\frac{\omega_0}{2}+\varepsilon_1^{-}}|\varphi_1^{-}\rangle\right)
\end{eqnarray}

\begin{eqnarray}
|\Psi_{g,0(\alpha^2)}^{(2)}\rangle=&&-\lambda^2\alpha^2\Bigg\{\frac{1}{\frac{\omega_0}{2}+\varepsilon_1^{+}}\Big(\frac{cos\theta_0(sin\theta_0sin\theta_1-\sqrt{2}cos\theta_0cos\theta_1)}{\frac{\omega_0}{2}+\varepsilon_0^{+}}
\nonumber\\&&-\frac{sin\theta_0(cos\theta_0sin\theta_1+\sqrt{2}sin\theta_0cos\theta_1)}{\frac{\omega_0}{2}+\varepsilon_0^{-}}\Big)|\varphi_1^{+}\rangle
\nonumber\\&&+\frac{1}{\frac{\omega_0}{2}+\varepsilon_1^{-}}\Big(\frac{cos\theta_0(sin\theta_0cos\theta_1+\sqrt{2}cos\theta_0sin\theta_1)}{\frac{\omega_0}{2}+\varepsilon_0^{+}}
\nonumber\\&&-\frac{sin\theta_0(cos\theta_0cos\theta_1-\sqrt{2}sin\theta_0sin\theta_1)}{\frac{\omega_0}{2}+\varepsilon_0^{-}}\Big)|\varphi_1^{-}\rangle
\nonumber\\&&+\frac{1}{2}\left[\left(\frac{cos\theta_0}{\frac{\omega_0}{2}+\varepsilon_0^{+}}\right)^2+\left(\frac{sin\theta_0}{\frac{\omega_0}{2}+\varepsilon_0^{-}}\right)^2\right]|g,0\rangle\Bigg\}
\end{eqnarray}

\begin{eqnarray}
|\Psi_{g,0(CRT)}^{(2)}\rangle=&&-\lambda^2\Bigg\{-\sqrt{3}sin\theta_1cos\theta_1\bigg[\frac{sin\theta_3}{\frac{\omega_0}{2}+\varepsilon_3^{+}}\left(\frac{1}{\frac{\omega_0}{2}+\varepsilon_1^{+}}-\frac{1}{\frac{\omega_0}{2}+\varepsilon_1^{-}}\right)|\varphi_3^{+}\rangle
\nonumber\\&&+\frac{cos\theta_3}{\frac{\omega_0}{2}+\varepsilon_3^{-}}\left(\frac{1}{\frac{\omega_0}{2}+\varepsilon_1^{+}}-\frac{1}{\frac{\omega_0}{2}+\varepsilon_1^{-}}\right)|\varphi_3^{-}\rangle\bigg]
\nonumber\\&&+\frac{1}{2}\left[\left(\frac{sin\theta_1}{\frac{\omega_0}{2}+\varepsilon_1^{+}}\right)^2+\left(\frac{cos\theta_1}{\frac{\omega_0}{2}+\varepsilon_1^{-}}\right)^2\right]|g,0\rangle\Bigg\}
\end{eqnarray}

\begin{eqnarray}
|\Psi_{g,0(\alpha,CRT)}^{(2)}\rangle=&&-\lambda^2\alpha\Bigg\{-\frac{1}{\frac{\omega_0}{2}+\varepsilon_0^{+}}\bigg[\frac{sin\theta_1}{\frac{\omega_0}{2}+\varepsilon_1^{+}}\left(sin\theta_1sin\theta_0-\sqrt{2}cos\theta_1cos\theta_0\right)
\nonumber\\&&+\frac{cos\theta_1}{\frac{\omega_0}{2}+\varepsilon_1^{-}}\left(cos\theta_1sin\theta_0+\sqrt{2}sin\theta_1cos\theta_0\right)\bigg]|\varphi_0^{+}\rangle
\nonumber\\&&-\frac{1}{\frac{\omega_0}{2}+\varepsilon_0^{-}}\bigg[\frac{sin\theta_1}{\frac{\omega_0}{2}+\varepsilon_1^{+}}\left(sin\theta_1cos\theta_0+\sqrt{2}cos\theta_1sin\theta_0\right)
\nonumber\\&&+\frac{cos\theta_1}{\frac{\omega_0}{2}+\varepsilon_1^{-}}\left(cos\theta_1cos\theta_0-\sqrt{2}sin\theta_1sin\theta_0\right)\bigg]|\varphi_0^{-}\rangle
\nonumber\\&&+\frac{1}{\frac{\omega_0}{2}+\varepsilon_2^{+}}\Big[\sqrt{2}sin\theta_2\left(\frac{cos^2\theta_0}{\frac{\omega_0}{2}+\varepsilon_0^{+}}+\frac{sin^2\theta_0}{\frac{\omega_0}{2}+\varepsilon_0^{-}}\right)
\nonumber\\&&-\bigg(\frac{sin\theta_1(\sqrt{2}sin\theta_1sin\theta_2-\sqrt{3}cos\theta_1cos\theta_2)}{\frac{\omega_0}{2}+\varepsilon_1^{+}}
\nonumber\\&&+\frac{cos\theta_1(\sqrt{2}cos\theta_1sin\theta_2+\sqrt{3}sin\theta_1cos\theta_2)}{\frac{\omega_0}{2}+\varepsilon_1^{-}}\bigg)\Big]|\varphi_2^{+}\rangle
\nonumber\\&&+\frac{1}{\frac{\omega_0}{2}+\varepsilon_2^{-}}\Big[\sqrt{2}cos\theta_2\left(\frac{cos^2\theta_0}{\frac{\omega_0}{2}+\varepsilon_0^{+}}+\frac{sin^2\theta_0}{\frac{\omega_0}{2}+\varepsilon_0^{-}}\right)
\nonumber\\&&-\bigg(\frac{sin\theta_1(\sqrt{2}sin\theta_1cos\theta_2+\sqrt{3}cos\theta_1sin\theta_2)}{\frac{\omega_0}{2}+\varepsilon_1^{+}}
\nonumber\\&&+\frac{cos\theta_1(\sqrt{2}cos\theta_1cos\theta_2-\sqrt{3}sin\theta_1sin\theta_2)}{\frac{\omega_0}{2}+\varepsilon_1^{-}}\bigg)\Big]|\varphi_2^{-}\rangle\Bigg\}
\end{eqnarray}

\section{the wave function of the higher-order dressed state}
The first-order wave function correction of the higher order dressed state is given by: $|\Psi_{k}^{\pm(1)}\rangle=|\Psi_{k(\alpha)}^{\pm(1)}\rangle+|\Psi_{k(CRT)}^{\pm(1)}\rangle$, where
\begin{eqnarray}
&&|\Psi_{k(\alpha)}^{+(1)}\rangle=-\lambda\alpha\Bigg\{\sum_{m\neq k}\frac{1}{\varepsilon_k^{+}-\varepsilon_m^{+}}\Bigg[sin\theta_m sin\theta_k\left(\sqrt{k+1}\langle m|k+1\rangle+\sqrt{k}\langle m|k-1\rangle\right)
\nonumber\\&&-cos\theta_m cos\theta_k\left(\sqrt{k+2}\langle m+1|k+2\rangle+\sqrt{k+1}\langle m+1|k\rangle\right)\Bigg]|\varphi_m^{+}\rangle
\nonumber\\&&+\sum_{m}\frac{1}{\varepsilon_k^{+}-\varepsilon_m^{-}}\Bigg[cos\theta_m sin\theta_k\left(\sqrt{k+1}\langle m|k+1\rangle+\sqrt{k}\langle m|k-1\rangle\right)
\nonumber\\&&+sin\theta_m cos\theta_k\left(\sqrt{k+2}\langle m+1|k+2\rangle+\sqrt{k+1}\langle m+1|k\rangle\right)\Bigg]|\varphi_m^{-}\rangle
\nonumber\\&&-\frac{1}{\varepsilon_k^{+}+\frac{\omega_0}{2}}\sqrt{k+1}cos\theta_k\langle0|k\rangle|g,0\rangle\Bigg\}\\
\nonumber\\&&|\Psi_{k(CRT)}^{+(1)}\rangle= -\lambda\Bigg\{\sum_{m\neq k}\frac{1}{\varepsilon_k^{+}-\varepsilon_m^{+}}\Bigg[\sqrt{k+2}sin\theta_m cos\theta_k\langle m|k+2\rangle
\nonumber\\&&+\sqrt{k}cos\theta_m sin\theta_k\langle m+1|k-1\rangle\Bigg]|\varphi_m^{+}\rangle
\nonumber\\&&+\sum_{m}\frac{1}{\varepsilon_k^{+}-\varepsilon_m^{-}}\Bigg[\sqrt{k+2}cos\theta_m cos\theta_k\langle m|k+2\rangle
-\sqrt{k}sin\theta_m sin\theta_k\langle m+1|k-1\rangle\Bigg]|\varphi_m^{-}\rangle
\nonumber\\&&+\frac{1}{\varepsilon_k^{+}+\frac{\omega_0}{2}}\sqrt{k}sin\theta_k\langle0|k-1\rangle|g,0\rangle\Bigg\}\\
&&|\Psi_{k(\alpha)}^{-(1)}\rangle=-\lambda\alpha\Bigg\{\sum_{m}\frac{1}{\varepsilon_k^{-}-\varepsilon_m^{+}}\Bigg[sin\theta_m cos\theta_k\left(\sqrt{k+1}\langle m|k+1\rangle+\sqrt{k}\langle m|k-1\rangle\right)
\nonumber\\&&+cos\theta_m sin\theta_k\left(\sqrt{k+2}\langle m+1|k+2\rangle+\sqrt{k+1}\langle m+1|k\rangle\right)\Bigg]|\varphi_m^{+}\rangle
\nonumber\\&&+\sum_{m\neq k}\frac{1}{\varepsilon_k^{-}-\varepsilon_m^{-}}\Bigg[cos\theta_m cos\theta_k\left(\sqrt{k+1}\langle m|k+1\rangle+\sqrt{k}\langle m|k-1\rangle\right)
\nonumber\\&&-sin\theta_m sin\theta_k\left(\sqrt{k+2}\langle m+1|k+2\rangle+\sqrt{k+1}\langle m+1|k\rangle\right)\Bigg]|\varphi_m^{-}\rangle
\nonumber\\&&+\frac{1}{\varepsilon_k^{-}+\frac{\omega_0}{2}}\sqrt{k+1}sin\theta_k\langle0|k\rangle|g,0\rangle\Bigg\}\\
\nonumber\\&&|\Psi_{k(CRT)}^{-(1)}\rangle= -\lambda\Bigg\{\sum_{m}\frac{1}{\varepsilon_k^{-}-\varepsilon_m^{+}}\Bigg[-\sqrt{k+2}sin\theta_m sin\theta_k\langle m|k+2\rangle
\nonumber\\&&+\sqrt{k}cos\theta_m cos\theta_k\langle m+1|k-1\rangle\Bigg]|\varphi_m^{+}\rangle
\nonumber\\&&-\sum_{m\neq k}\frac{1}{\varepsilon_k^{-}-\varepsilon_m^{-}}\Bigg[\sqrt{k+2}cos\theta_m sin\theta_k\langle m|k+2\rangle
+\sqrt{k}sin\theta_m cos\theta_k\langle m+1|k-1\rangle\Bigg]|\varphi_m^{-}\rangle
\nonumber\\&&+\frac{1}{\varepsilon_k^{-}+\frac{\omega_0}{2}}\sqrt{k}cos\theta_k\langle0|k-1\rangle|g,0\rangle\Bigg\}
\end{eqnarray}

\end{appendix}

\end{document}